\documentclass[prd,aps,reprint,amsmath,amssymb,onecolumn,nofootinbib,superscriptaddress]{revtex4-2}

\usepackage{wasysym,graphicx}
\usepackage{epstopdf}
\usepackage{alphalph}

\usepackage{xcolor}
\usepackage[colorlinks,linkcolor=blue]{hyperref}
\usepackage[normalem]{ulem}

\usepackage{orcidlink}
\usepackage{academicons}

\definecolor{orcidlogocol}{HTML}{A6CE39}
\newcommand{\orcid}[1]{\href{https://orcid.org/#1}{\textcolor[HTML]{A6CE39}{\aiOrcid}}}
\allowdisplaybreaks

\def\al{\alpha}
\def\be{\beta}
\def\ga{\gamma}
\def\de{\delta}
\def\ep{\epsilon}

\def\et{\eta}

\def\ka{\kappa}
\def\la{\lambda}
\def\rh{\rho}

\def\ta{\tau}

\def\ph{\phi}

\def\om{\omega}

\def\Si{\Sigma}

\def\Ph{\Phi}
\def\Ps{\Psi}
\def\Om{\Omega}

\def\mn{{\mu\nu}}

\def\abgd{{\al\be\ga\de}}

\def\prt{\partial}

\def\pt#1{\phantom{#1}}

\newcommand{\beq}{\begin{equation}}
\newcommand{\eeq}{\end{equation}}
\newcommand{\bal}{\begin{aligned}}
\newcommand{\eal}{\end{aligned}}

\newcommand{\rf}[1]{(\ref{#1})}

\def\sb{\overline{s}}
\def\htw{\tilde{h}}
\def\stw{\tilde{s}}
\def\Sitw{\tilde {\Si}}

\def\cL{{\cal L}}

\def\rf#1{(\ref{#1})}

\begin{document}

\title{You shall not pass! - explicit diffeomorphism violation ``no-go" constraints and discontinuities}

\author{Quentin G. Bailey\,\orcidlink{0000-0001-5917-6850}}
\email{baileyq@erau.edu}
\author{Kellie O'Neal-Ault\,\orcidlink{0000-0002-6645-4473}}
\email{aultk@erau.edu}
\affiliation{Embry-Riddle Aeronautical University, Prescott, AZ, 86301, USA}

\author{Nils A. Nilsson\,\orcidlink{0000-0001-6949-3956}}
\email{nilsson@ibs.re.kr}
\affiliation{Cosmology, Gravity and Astroparticle Physics Group, Center for Theoretical Physics of the Universe,
Institute for Basic Science, Daejeon 34126, Korea}
\affiliation{SYRTE, Observatoire de Paris, Université PSL, CNRS, LNE, Sorbonne Universit\'e, 61 avenue de l’Observatoire, 75014 Paris, France}

\date{\today}

\begin{abstract}
This paper collects several results in the study of the explicit symmetry-breaking limit of the effective-field theory (EFT) description of diffeomorphism and local Lorentz-symmetry breaking, where we generalize a subset of the EFT framework (the ``minimal" sector). 
It is well known that no-go constraints may arise in cases of explicit symmetry-breaking in curved spacetime as a consequence of the Bianchi identities; we show in this work that certain terms in the action can be countenanced and used to cancel would-be no-go constraints, 
at least in the linearized gravity limit.
Nonetheless, 
we go on to find more potential issues, 
and we show that one particular explicit breaking subset of the EFT,
while evading direct no-go constraints, 
results in a discontinuity - unsuppressed additional polarizations for gravitational waves.
In a general treatment of the explicit breaking EFT, 
but confined to linearized gravity, 
we explicitly show the existence of an extra degree of freedom, 
independent of coordinates. 
We find extra polarizations of gravitational waves
in the solutions, 
with a scalar mode unsuppressed by any coefficient,
which could render these cases ruled out by observations.
\end{abstract}

\maketitle

\section{Introduction}\label{sec:intro}

The notion that the spacetime symmetries underlying General Relativity (GR) and the Standard Model (SM) might not be exact at high energy scales has been considered as a possible signature of new physics for decades
\cite{ksstring89,kp95,gp99,chkl01}. 
Presently, 
the interest in searching for such departures from known physics is high, 
both at the theoretical and phenomenological level 
\cite{Mariz:2022oib,Addazi:2021xuf,safronova18,Will:2014kxa,Bailey:2023pfd,yunes16}. 
It is generally expected that GR and the SM are not the fundamental descriptions of nature, 
but instead arise in the low-energy limit of some effective-field theory (EFT) which accurately describes the physics at presently attainable energy scales \cite{Weinberg:2009bg}.
This point of view is motivated by the expectation of a single unified theory encompassing all fundamental interactions, which includes an as yet unknown quantum theory of gravity of which GR is the low-energy limit. 
To experimentally probe these ideas,
one posits higher-order corrections in the form of spacetime-symmetry breaking operators (such as local Lorentz symmetry and diffeomorphism symmetry) appearing as correction terms in the low-energy EFT \cite{kp95}. 
The search for such signals has generated a large number of stringent constraints across a vast number of experiments and observations \cite{datatables}. 

The origin of any hypothetical breaking of local Lorentz and diffeomorphism symmetry can affect its phenomenological signatures; 
generally, 
broken diffeomorphism and/or broken Lorentz symmetry is signaled by the presence of one or more background tensor fields ($t_{\mu\nu...}$) in spacetime that couple to existing matter or fields.
Using this notion, 
a theory agnostic general EFT framework for tests of these symmetries has been in use for decades \cite{ck97,ck98,k04}.
In this framework, 
the origin of the Lorentz breaking terms, 
which generically involve indexed coefficients contracted with field operators, 
is largely not specified.
Exceptions occur when curved spacetime is involved:
there, 
to solve field equations and calculate observables, 
one generally has to specify whether the symmetry breaking is explicit ({\it a priori} given backgrounds $t_{\mu\nu\hdots}$) or spontaneous, 
where in the latter case the backgrounds are vacuum expectation values of dynamical fields ($t_{\mu\nu\hdots}\to\langle t\rangle_{\mu\nu\hdots}$);
attention has been given to both cases in the literature.
It was shown in Ref.\ \cite{bk04} and elsewhere 
that conflicts between curved geometry
and explicit breaking backgrounds can arise.
More recently, 
attempts to circumvent these issues, 
and find solutions for special cases like cosmology, 
have been countenanced \cite{Bonder:2020fpn, abn21, Reyes:2021cpx}.
Alternatively, the geometry of spacetime may not even be Riemannian at all 
\cite{kl21nR}, 
or could be a Riemann-Finsler geometry \cite{Kostelecky:2011qz,Lammerzahl:2012kw}, which in some ways can be considered more natural when spacetime symmetries are broken.

On the other hand, 
the {\it spontaneous}-symmetry breaking case is attractive for its elegance
and its avoidance of direct no-go constraints affecting the explicit case.
Furthermore,
with the spontaneous breaking case, 
one has a full set of field equations one can actually attempt to solve \cite{ks89bb,Jacobson:2000xp,Kostelecky:2009zr,abk10},
thus allowing for study of the classical and quantum field theory \cite{Kostelecky:2000mm,Jacobson:2000xp,bk05,Eling:2006ec,Kostelecky:2009zr,Carroll:2009em,abk10,casana18,Delhom:2022xfo,Liang:2022hxd,Mai:2024lgk}.
Nonetheless, 
spontaneous breaking it is not without its own challenges, 
such as unobserved Nambu-Goldstone modes
\cite{bk05,Carroll:2009mr,bk08}.
In addition, 
the spontaneous symmetry breaking requires a potential energy function that admits a nonzero minimum and possible massive modes \cite{bk08}, 
the latter of which can result in the unboundedness of the Hamiltonian \cite{Bluhm:2008yt,Carroll:2009em},
singular Hamiltonians \cite{Seifert:2019kuz}
and possible ambiguities \cite{Bonder:2015jra, pottATT}.

In this paper, 
we will revisit a relatively simple subset of the EFT in the explicit breaking limit, 
partly as an extension of the Hamiltonian results in Ref.\ \cite{abn21}, 
but using a Lagrangian-based approach.
We show that certain combinations of traced coefficient terms in the action can be chosen to cancel some no-go terms in the constraint equations resulting from the geometric Bianchi identities.
Even so, 
we go to show that other problems arise, such as the lack of a continuous limit to GR (discontinuities), by studying a special case that is ultimately ruled out by experiment.
In addition we perform a detailed study of the linearized gravity limit with explicit-symmetry breaking terms, 
detailing the degrees of freedom occurring in this case, 
and pointing out where discontinuities arise, 
and the impact on observation and experiment.

The paper is organized as follows: In Section~\ref{sec:minimaltrace}, 
we write down a generalized version of the gravitational EFT and explore the constraints on the stress-energy tensor.
In Section~\ref{sec:simplest} we investigate a subset of this construction with the simplest trace corrections, where we including a match to scalar-tensor theory and show the appearance of an extra unsuppressed gravitational-wave mode. 
In Section~\ref{sec:genlin} we study the linearized gravity limit and show the appearance of extra degrees of freedom using, 
in part, 
a momentum-space solution, 
and we categorize the propagating/non-propagating nature of the extra modes. 
Finally, in Section~\ref{sec:summary} we summarize the work and discuss implications of the results.

The notations and conventions in this work match other papers by the authors \cite{bk06,bkx15,abn21,Bailey:2023lzy}.
In particular, 
we use the spacetime metric signature $(-+++)$.
Spacetime indices on tensor components are Greek letters ($\mu$, $\nu$, $\hdots$), 
while spatial indices are Latin ($i$, $j$, $\hdots$).
We denote the covariant derivative as $\nabla_\mu$, and use $\nabla^2$ as the Laplacian operator in some places.

\section{The minimal gravitational EFT with trace corrections}\label{sec:minimaltrace}

Symmetry-breaking terms in the EFT gravitational action are constructed
with coefficients $k_{\al\be...}$ contracted with the Riemann curvature $R_{\ka\la\mu\nu}$ and covariant derivative 
terms $\nabla_\ep ... R_{\ka\la\mu\nu}$ with increasing mass dimension $d$.
The general EFT action for the gravity sector for arbitrary mass dimension terms has been discussed elsewhere
\cite{k04,bkx15,km16,kl21}.
Here we focus on the $d=4$ case; however, unlike previous treatments, 
we allow also for independent terms with various traces of coefficients that involve the spacetime metric $g_\mn$.
In certain cases these terms can contribute differently to the equations of motion;
for instance, 
this is clear if one compares the field equations for the Lagrange density $\cL \sim s_\mn R^\mn$, 
with symmetric two-tensor coefficients $s_\mn$,
to a Lagrange density term $\cL \sim k^\al_{~\mu\al\nu} R^\mn$, 
with contracted coefficients $k^\al_{~\mu\al\nu}$. 
The presence of the metric in the latter, 
due to the contraction, 
affects the variation with respect to $g_\mn$, 
spoiling the naive tactic of absorbing $k^\al_{~\mu\al\nu}$ into $s_\mn$.

Thus, 
considering couplings to Riemann curvature only, 
we have,
\begin{equation}
    S = \frac{1}{2\kappa}\int d^4x \sqrt{-g}\left[R\left(1-e_1 u+e_2 s^\alpha_{~\alpha}+e_3 t^{\alpha\beta}_{~~\alpha\beta}\right)+R^{\mu\nu}\left(e_4s_{\mu\nu}+e_5t_{\mu~\nu\alpha}^{~\alpha}\right)+e_6t_{\alpha\beta\mu\nu}R^{\alpha\beta\mu\nu}\right]+S_M,
    \label{genAct}
\end{equation}
where $\kappa = 8\pi G_N$, 
$R^{\mu\nu}$ is the Ricci tensor and $R$ is the curvature scalar.
Also, 
$e_{1\hdots6}$ are constants, 
and the subset containing $e_1$, $e_4$, and $e_6$ makes up what is usually known as the ``minimal" subset of the gravity EFT \cite{k04}, 
where we have added the trace corrections $e_2$, $e_3$, and $e_5$, 
which generalizes the minimal EFT sector. 
Note that $s_\mn$ and $t_{\al\be\ga\de}$ are not assumed traceless here. 
Since we are considering explicit breaking, 
we do not include dynamical terms in the action for the coefficients \cite{b21}.
We also neglect any mass-type terms for the gravitational field, 
such as those considered in massive gravity models \cite{PhysRevD.82.044020,Kostelecky:2021xhb} or such terms in the EFT framework \cite{km18}.
The case where only $e_1 \neq 0$ and/or $e_4 \neq 0$ with explicit diffeomorphism violation has been studied with the Dirac-Hamiltonian method in references \cite{abn21,Reyes:2021cpx,Reyes:2023sgk};
we complement this work with a Lagrangian-based analysis.

Note here that explicit diffeomorphism violation can be seen under the ``particle" (infinitesimal) diffeomorphism transformation where the coefficients $u$, $s_\mn$, and $t_{\alpha\beta\mu\nu}$ do not transform ($s_{\mu\nu}\overset{\rm part}{\longrightarrow} s_{\mu\nu}$) while the metric $g_\mn$ transforms as a two-tensor ($g_{\mu\nu}\overset{\rm part}{\longrightarrow} g_{\mu\nu} + \mathcal{L}_{\xi}g_{\mu\nu}$) \cite{bluhm15}; conversely, 
under a general coordinate transformation both $s_\mn$ and $g_\mn$ transform as two-tensors and the action \rf{genAct} remains coordinate covariant.

The field equations are obtained from the action with a variation with respect to $g_\mn$ \cite{Reyes:2023sgk}, 
while holding fixed all the indexed coefficients in the ``downstairs", 
co-vector position ($s_\mn,~t_{\al\be\ga\de}$). 
This is a particular choice of background, 
as physical results can differ for other choices \cite{kl21}.
Note in particular,
\beq
\bal
s^\mu_{\pt{\mu}\mu} &= g^\mn s_\mn,\\
t^\al_{~\mu\al\nu}
&= g^{\al\be} t_{\al\mu\be\nu},
\label{strace}
\eal
\eeq
so that $s_\mn$ and $t_{\al\mu\be\nu}$ (in the lowered index position) are held fixed for the variation.
This implies that the coupling with $s^\al_{~\al}$
in \rf{genAct} is not merely a scalar field, 
like in scalar-tensor models.
We obtain from this variation,
\begin{equation}\label{eq:generalEE}
    -G^\mn+\sum_{n=1}^6 e_n E_{(n)}^\mn 
    + \ka T_M^\mn = 0,
\end{equation}
where the tensors $E_{(n)}^\mn$ are defined as
\begin{equation}
\begin{aligned}
    E_{(1)}^\mn &= G^\mn u+\left(g^\mn \nabla^2-\nabla^\mu \nabla^\nu \right)u \\
    E_{(2)}^\mn &= -R s^\mn-G^\mn s^\al_{~\al}-\left(g^\mn \nabla^2-\nabla^\mu \nabla^\nu \right) s^\al_{~\al} \\
    E_{(3)}^\mn &= -2R t^{\mu\al\nu}_{~~~~\al}-G^\mn t^{\al\be}_{~~\al\be}-
    \left(g^\mn \nabla^2-\nabla^\mu \nabla^\nu \right)
    t^{\al\be}_{~~\al\be}.
\end{aligned}
\end{equation}
The tensors $E_{(4)\hdots(6)}$ are lengthy and we display them in Appendix~\ref{app:fullexpr}.
In the absence of any symmetry breaking the $E_{(n)}^\mn$'s all vanish.

One place where one runs into challenges in the explicit-symmetry breaking scenario is via the conservation laws implied by the covariant divergence of the field equations \rf{eq:generalEE}.
These must be satisfied in order for the equations to be consistent.
By taking into account geometric identities,
that is, 
the traced Bianchi identities $\nabla_\mu G^\mn = 0$, 
and 
assuming conservation of the matter energy-momentum tensor
$\nabla_\mu T_M^\mn=0$, 
we find the covariant divergence of Eq.~\rf{eq:generalEE} to be
\begin{equation}\label{eq:generalBI}
    \sum_{n=1}^6 e_n\nabla_\mu E_{(n)}^\mn = 0,
\end{equation}
where 
\begin{equation}
    \begin{aligned}
        \nabla_\mu E_{(1)}^\mn &= -\frac{1}{2}R\nabla^\nu u \\
        \nabla_\mu E_{(2)}^\mn &= \frac{1}{2}R\nabla^\nu s^\alpha_{~\alpha}-\nabla_\mu \left(s^\mn R\right) \\
        \nabla_\mu E_{(3)}^\mn &= \frac{1}{2}R\nabla^\nu t^{\alpha\be}_{~~~\alpha\be}-2\nabla_\mu \left(t^{\mu\al\be}_{~~~~\al}R\right),
        \label{divterms}
    \end{aligned}
\end{equation}
and the rest are displayed in Appendix~\ref{app:fullexpr}. The Bianchi identities place severe constraints on explicit-breaking models, but previous work has shown that they can be evaded in certain scenarios; see for example \cite{ONeal-Ault:2021uwu, Reyes:2022dil}, and more recently \cite{Reyes:2024ywe}.

The primary feature we emphasize here is that various terms that show up in Eq.~\rf{eq:generalBI} for each coupling $e_n$, 
starting from $e_1$, 
can in principle be cancelled by successive $e_{n+1}$ terms.
For example, 
consider from Eq.~\rf{DEterms} the $e_4$ contribution
$\nabla_\mu E_{(4)}^{\mu\nu} = \tfrac 12 R^{\alpha\be}\nabla^\nu s_{\alpha\be}
-\tfrac 12 s^{\alpha\nu}\nabla_\alpha R
-R^{\alpha\be}\nabla_\be s^{\nu}_{~\alpha}, $.
All three of these terms can be canceled by corresponding terms in 
$\nabla_\mu E_{(5)}^{\mu\nu} =
- \tfrac{1}{2} t^{\al \be \nu }{}_{\be } \nabla_{\al }R 
+ \tfrac{1}{2} R^{\alpha \be } \nabla^{\nu }t_{\al \ga \be }{}^{\ga} 
- R^{\alpha \be } \nabla_{\al }t_{\be \ga }{}^{\nu \ga }+...$, 
by choosing $e_5 t_{\mu\al\nu}^{~~~~~\al}=-e_4 s_\mn$.
This does not eliminate all the terms in $\nabla_\alpha E_{(5)}^{\alpha\beta}$, 
however. One might be tempted to use this constraint to reduce the equations, but one would then ruin the notion of explicit breaking. As a result, variations of $s_{\mu\nu}$ with respect to the inverse metric would not give zero, and $s_{\mu\nu}$ would no longer be unaffected by particle diffeomorphisms.
Morever, 
this substitution at the level of the action would change the field equations;
additionally the metric inverse is $g^\mn$ shows up and thus this relation equates dynamical variables to prescribed, nondynamical background coefficients.
Henceforth in this paper we set $e_5=0=e_6$ for simplicity;
nonetheless, some interesting features arise from this sort of balance as we show below.

We consider now the special case of Eq.~\rf{genAct} with a two tensor $s_\mn$ as well as the scalar $u$ and the saturated trace $t^{\alpha\beta}_{~~\alpha\beta}$.
The field equations from Eq.~\rf{eq:generalEE} in this case are
\begin{equation}\label{eq:Einstein}
\begin{aligned}
    &&-G^{\mu\nu}+e_1\Big[G^{\mu\nu}u-\nabla^\nu\nabla^\mu u+g^{\mu\nu}\nabla_\alpha\nabla^\alpha u\Big]+e_2 \Big[-G^{\mu\nu}s^\alpha_{~\alpha} - R s^{\mu\nu} -g^{\mu\nu} \nabla_\beta\nabla^\beta s^\alpha_{~\alpha} + \nabla^\nu\nabla^\mu s^\alpha_{~\alpha}\Big] \\&&
    +e_3\Big[\nabla^\nu\nabla^\mu t^{\alpha\beta}_{~~\alpha\beta}-2Rt^{\mu\alpha\nu}_{~~~~\alpha}-G^{\mu\nu}t^{\alpha\beta}_{~~\alpha\beta}-g^{\mu\nu}\nabla_\lambda\nabla^\lambda t^{\alpha\beta}_{~~\alpha\beta}\Big]+e_4\Big[\tfrac{1}{2}g^{\mu\nu}G^{\alpha\beta}s_{\alpha\beta}+\tfrac{1}{4}g^{\mu\nu} R s^\alpha_{~\alpha} - 2 G^{(\mu \alpha} s^{\nu )}_{~\alpha}\\&&-\tfrac{1}{2}\nabla_\alpha\nabla^\alpha s^{\mu\nu}+\nabla_\alpha\nabla^{( \mu } s^{\nu )\alpha} - \tfrac{1}{2}g^{\mu\nu}\nabla_\alpha\nabla_\beta s^{\alpha\beta}-Rs^{\mu\nu}\Big]+\kappa T^{\mu\nu} = 0
\end{aligned}
\end{equation}
where $G^{\mu\nu} = R^{\mu\nu}-\tfrac{1}{2}g^{\mu\nu}R$ is the Einstein tensor.
Note that a significant number of the terms are tensorially similar.
From Eq.~\rf{eq:generalBI}, 
the constraints implied by the traced Bianchi identities yield
\begin{equation}
\begin{aligned}
   && -\tfrac{1}{2}e_1R\nabla^\nu u + e_2\Big[-R\nabla_\mu s^{\mu\nu}-s^\nu_{~\mu}\nabla^\mu R + \tfrac{1}{2}R\nabla^\nu s^\alpha_{~\alpha}\Big]
   + e_3\Big[2R\nabla_\alpha t^{\nu\beta~\alpha}_{~~\beta}+\tfrac{1}{2}R\nabla^\mu t^{\alpha\beta}_{~~\alpha\beta}-2t^{\nu\beta}_{~~\alpha\beta}\nabla^\alpha R\Big] \\ &&
   +e_4\Big[-G^{\mu\alpha}\nabla_\alpha s^\nu_{~\mu}-\tfrac{1}{2}R\nabla_\mu s^{\mu\nu} -\tfrac{1}{2}s^\nu_{~\mu}\nabla^\mu R +\tfrac{1}{2}G^{\mu\alpha}\nabla^\nu s_{\mu\alpha}+\tfrac{1}{4}R\nabla^\nu s^\alpha_{~\alpha}\Big] + \kappa\nabla_\mu T^{\mu\nu} = 0.
   \label{BI1}
    \end{aligned}
\end{equation}
These four equations must be satisfied by consistency;
if they are not satisfied,
this limit of the framework is deemed no-go due to the clash with Riemannian geometry.
Assuming that the matter stress energy tensor is separately conserved, 
and using the similarity of many of the terms, 
these four equations can be re-arranged as
\beq
\bal
&&-R [ \la_1 \nabla_\mu s^{\mu\nu}+2 e_3 \nabla_\mu t^{\mu\al\nu}_{~~~~\al} ]
- [\la_1 s^{\nu\mu} + 2 e_3 t^{\nu\al\mu}_{~~~~\al}] \nabla_\mu R  
+\tfrac 12 R \nabla^\nu [-e_1 u + \la_1 s^\al_{~\al} + 
e_3 t^{\al\be}_{~~\al\be}]
\\
&&
-e_4 \tfrac{1}{2}g^{\nu\rh}G^{\mu\al}\Big(\nabla_\mu s_{\rh\al}+\nabla_\al s_{\rh\mu}-\nabla_\rh s_{\mu\al}\Big) = 0,
\label{BI2}
\eal
\eeq
where $\la_1 = e_2+\tfrac 12 e_4$.
A striking result of \rf{BI2} is that one can actually propose relations among the constants $e_i$, and 
the background coefficients,
that satisfy this constraint, 
which can be thought of as a kind of ``detailed balance".
 Different choices of coefficient values represent distinct models, for example:
\begin{itemize}
\item[({\it i})]
$e_4=0$,
$\la_1 s^\mn = -2e_3 t^{\mu\al\nu}_{~~~~\al}$, and $e_1 u= \la_1  s^\al_{~\al} + e_3 t^{\al\be}_{~~\al\be}$; the field equations reduce to GR, but produces non-viable constraints on the dynamical metric.
\item[({\it ii})]
$e_4=0$; 
a subset of the constraints are eliminated (we explore this case in the next section).
\item[({\it iii})]
$e_1=0=e_3$, 
$\la_1=e_2 + \tfrac 12 e_4=0$;
the linearized limit satisfies the constraints. 
\end{itemize}
The choice ({\it iii}) shows that it is possible for the effects of Lorentz violation to evade linearized no-go constraints in this limit, 
providing an exception to those posited for generic explicit breaking terms in \cite{Bluhm:2019ato,kl21}.

As a warm-up example, we choose the subcase (iii) when $e_1=e_3=0$, take the constraint equations in Eq.~(\ref{BI2}) and impose the detailed-balance condition $e_2+\tfrac{1}{2}e_4=0$. 
We find,
\begin{equation}
g^{\nu\rho}G^{\mu\alpha}\left(\nabla_\mu s_{\rho\alpha}+\nabla_\alpha s_{\rho\mu}-\nabla_\rho s_{\mu\alpha}\right) = 0.
\label{constraintsIII}
\end{equation}
These four equations represent local constraints on a nonlinear combination of the background and the curvature.
We do not attempt to solve these constraints in generality here but we consider that one might use them to impose constraints on the background $s_\mn$ itself; we grant that this is a philosophical departure from the concept of an {\it a priori} independent background \cite{abn21,Reyes:2021cpx}.
Thus, the constraints could be solved if we have
\begin{equation}\label{eq:ds0}
    \nabla_\mu s_{\rho\alpha}+\nabla_\alpha s_{\rho\mu}-\nabla_\rho s_{\mu\alpha} = 0,
\end{equation}
from which we find
$\nabla_\mu s^\alpha_{~\alpha} = 0$. Therefore, in the subcase (\emph{iii}), the trace of the coefficients $s_{\mu\nu}$ must be constant in the chosen coordinates in order for the traced Bianchi identities to be satisfied.
We also notice that $R=0$ (without assuming (\emph{iii})) reduces Eq.~(\ref{BI2}) to
\begin{equation}
    e_4 \tfrac{1}{2}g^{\nu\rho}R^{\mu\alpha}\Big(\nabla_\mu s_{\rho\alpha}+\nabla_\alpha s_{\rho\mu}-\nabla_\rho s_{\mu\alpha}\Big) = 0,
\end{equation}
which is solved by the same trace constraint on $s_{\mu\nu}$ as in the previous example. Although we find the same covariant constraint equation for $s_{\mu\nu}$, the physical situation is entirely different, as we are now restricted to Ricci flat metrics. 

\section{Consequences of the trace-type terms in the action}
\label{sec:simplest}

In this section we consider a subset of the action \rf{genAct}, 
and study its consequences in the explicit symmetry-breaking limit.
Allowing $e_1-e_3$ 
to be nonzero, 
as in the special case
({\it ii}) in the previous section, 
the field equations are
\beq
-G^\mn (1+\Ph) 
-( e_2 s^\mn + 2 e_3 t^{\mu \al\nu}_{~~~~\al}) R 
- (g^\mn \nabla^2 -\nabla^\mu \nabla^\nu )\Ph + 8\pi G_N T^\mn = 0,
\label{FE}
\eeq
where we introduce the scalar combination
$\Ph=-e_1 u +e_2 s^\al_{~\al} +e_3 t^{\al\be}_{~~~\al\be}$.
The constraint equations \rf{BI2} 
become
\beq
-\nabla_\mu [(e_2 s^\mn+2e_3t^{\mu\al\nu}_{~~~~\al})R] 
+ \frac 12 R \nabla^\nu \Phi =0. 
\label{constraint}
\eeq
This can be viewed as four constraints on the four would-be gauge degrees of freedom in this broken diffeomorphism action \cite{bluhm15}.
However, 
in order for this action to have any hope of matching observations in the weak-field gravity limit,
we must satisfy the weak-field limit of this constraint,
\beq
-(e_2 s_\mn + 2e_3 t_{\mu\al\nu}^{~~~~\al} ) \prt^\mu R_L =0, 
\label{Lconstraint}
\eeq
where we assume an asymptotically flat spacetime in which $s_\mn$ tends to a constant background, 
and $R_L$ is the linearized Ricci tensor.
In this context of linearized gravity any would-be diffeomorphism gauge degrees of freedom vanish from the Ricci scalar \cite{bluhm15}, 
similar to other scenarios \cite{bk05}.
The only plausible way to satisfy \rf{Lconstraint} and be consistent with perturbative constraints in the weak-field limit \cite{kl21}, 
is to ``solve" the constraint with 
the condition $R=0$.
Thus we get a scalar constraint equation
on the dynamical variables
\beq
R=0,
\label{constraint2}
\eeq
which holds now for the full curvature (not linearized).
The field equations then become,
\beq
-G^\mn (1+\Ph) - (g^\mn \nabla^2-\nabla^\mu \nabla^\nu ) \Ph + 8\pi G_N T^\mn = 0.
\label{FE3}
\eeq

Tracing the field equation \rf{FE3} yields an equation:
\beq
\boxed{%
-3 \nabla^\mu \nabla_\mu \Ph + 8\pi G_N T^\al_{\pt{\al}\al}=0.
}
\label{wave1}
\eeq
This gives a wave equation for the trace 
$\Ph$.
Since there is no other constraint on $\Ph$, 
this represents an extra degree of freedom
in the model.
Note that in general $\Ph$ is not merely composed of the background {\it a priori} coefficients but involves components of the metric via the traces like \rf{strace}; 
thus the extra degree of freedom lies in the spacetime metric.
We examine aspects of this case in the subsections to follow.
Note that the propagation of the extra mode appears of the standard form, 
so it does not appear to be a ghost mode from this perspective; we discuss this further in Section \ref{solution for the curvature in the wave zone}.

\subsection{Match to scalar-tensor models}
\label{Match to scalar-tensor models}

The action of a general scalar-tensor theory can be written as
\begin{equation}
    S = \frac{1}{2\kappa}\int d^4x \sqrt{-g} \left(\phi R -\frac{\omega(\phi)}{\phi}\nabla^\mu\phi\nabla_\mu\phi-V(\phi)\right)+S_M,
\end{equation}
where $S_M$ is an arbitrary matter action
\cite{Will:2018bme}. 
By setting the coupling $\omega(\phi)$ to a constant and the potential $V(\phi)=0$, one obtains
the original Brans-Dicke theory \cite{Brans:1961sx}.
The field equations \rf{FE3} can be directly mapped to the field equations of scalar-tensor theory as we show below. 
For a general scalar-tensor model with no potential term $V(\ph)=0$, 
the field equations can be written \cite{Weinberg:1972kfs} as
\beq
\begin{aligned}
G^\mn &= \frac{8\pi G_N}{\ph} (T_M)^\mn + \frac{\om}{\ph^2} \left( \nabla^\mu \ph \nabla^\nu \ph - \frac 12 g^\mn (\nabla \ph)^2 \right)
+  \frac{1}{\ph} \left( \nabla^\mu \nabla^\nu \ph - g^\mn \nabla^2 \ph \right),
\\
\nabla^2 \ph &= \frac {8\pi G_N}{3+2\om} (T_M)^\al_{\pt{\al}\al},
\label{scalar}    
\end{aligned}
\eeq
where $\om$ is a free parameter such that the theory reduces to GR when $\om\rightarrow \infty$.

First we observe that comparing the first of equations \rf{scalar}
with Eq.~\rf{FE3}, 
the term proportional to $\om$ is absent in Eq.~\rf{FE3}.
Therefore we must take $\om=0$.
Manipulation of Eq.~\rf{scalar} then yields
\beq
\ph G^\mn - 8\pi G_N (T_M)^\mn 
- \left( \nabla^\mu \nabla^\nu \ph 
- g^\mn \nabla^2 \ph \right)=0.
\label{scalar2}
\eeq
If we posit $\ph=1+\Ph$, 
then \rf{scalar2} matches \rf{FE3}.
Also, in this limit of $\om \rightarrow 0$,
Eq.~\rf{scalar2} and the field equation for $\ph$ in Eq.~\rf{scalar} together imply $R=0$.

In summary, 
there is a match of the field equations \rf{FE} with the Bianchi-identity implied condition $R=0$ to scalar tensor theory when $\om=0$ and 
$\ph=1+\Ph$.
Note that the current limits from observational tests are on the order of $\om > 10^4$: specifically, by measuring the Shapiro time delay in the Solar System using the Cassini spacecraft, it was found that $\omega > 4\cdot 10^4$ \cite{Will:2014kxa,Bertotti:2003rm}, and through the orbital decay of the binary PSR J1738+0333 the constraint $\omega > 2.5 \cdot 10^4$ was obtained \cite{Freire:2012mg}. 
This therefore shows that the physics from Eqs~\rf{FE3}
and \rf{wave1} are ruled out by experiment. We note that this correspondence was first pointed out in \cite{Reyes:2022mvm} through the use of a scalar corresponding to the $e_1$ term in our action (\ref{genAct}) for the standard minimal EFT where all coefficients are traceless. We explain this  result further below with a GW scenario.

\subsection{Solution for the curvature in the wave zone}
\label{solution for the curvature in the wave zone}

We next explore the gravitational wave scenario in the weak field limit.
First note that the observables for the typical gravitational-wave measurement scenario are the six components of the Riemann tensor $R_{0j0k}$ in locally freely falling coordinates \cite{Will:2018bme}. 
Specializing to small fluctuations $h_\mn$ around approximately flat spacetime $g_\mn = \et_\mn + h_\mn$, 
the following identity holds:
\begin{equation}
\Box R_\abgd = \prt_{\be\de} R_{\al\ga}-\prt_{\be\ga} R_{\al\de}+\prt_{\al\ga} R_{\be\de}-\prt_{\al\de} R_{\be\ga},
\label{identity}
\end{equation}
where $\Box = \de^{ij} \prt_i \prt_j - \prt_0^2$.
With \rf{constraint2} the Ricci scalar vanishes, 
and so we may substitute $R_\mn \rightarrow G_\mn$ in \rf{identity}.

From Eq.~\rf{FE3}, 
we have 
\beq
G_\mn = \lambda (\nabla_\mu \nabla_\nu - g_\mn \nabla^2 ) \Ph
+ 8\pi G_N \lambda T^\mn,
\label{FE4}
\eeq
where $\lambda = 1/(1+\Ph)$.
In the weak-field gravity limit, 
since any fluctuations in $\Ph$ are governed by \rf{wave1} and sourced by matter stress-energy, we can (if we also ignore any homogeneous solutions) expect them to be of the same approximate size as the fluctuations $h_\mn$ in a weak-field expansion.
So we can take the $\lambda$ multiplying terms in \rf{FE4} to be a constant $\bar{\lambda} = 1/(1+ {\overline \Ph})$, 
where $\overline \Ph$ is a constant, 
zeroth order trace.
Thus we obtain the weak-field approximate equation
\beq
G_\mn = \bar{\lambda} (\prt_\mu \prt_\nu - \et_\mn \Box ) \Ph+ 8\pi G_N \bar{\lambda} T_\mn,
\label{FE3wf}
\eeq
where the curvature tensor is linearized.
It is interesting to note here that the contribution to the energy density $T_{00}$ on the right-hand side is $\bar{\la}\nabla^2 \Ph$, 
which does not appear to be bounded from below.
Nonetheless we continue our investigation to solve for the (measureable) curvature.

When inserting the expression \rf{FE3wf} into \rf{identity} the partial derivative terms acting on $s^\al_{\pt{\al}\al}$ vanish due to the tensor symmetries of the terms on the right hand side of \rf{identity}.
This leaves the field equations as
\beq
\Box R_\abgd = 8 \pi G_N \bar{\lambda}(\prt_{\be\de} T_{\al\ga}-\prt_{\be\ga} T_{\al\de}+\prt_{\al\ga} T_{\be\de}-\prt_{\al\de} T_{\be\ga} )
- \bar{\lambda} (\prt_{\be\de} \et_{\al\ga}-\prt_{\be\ga} \et_{\al\de}+\prt_{\al\ga} \et_{\be\de}-\prt_{\al\de} \et_{\be\ga} ) \Box \Ph.
\label{identity2}
\eeq
Now it is clear we can use the wave equation for $\Ph$ 
(Eq.~\rf{wave1})
 to express the right-hand side entirely in terms of the stress-energy tensor, 
whereupon we arrive at a wave equation for the curvature itself sourced by derivatives of the stress-energy tensor.
We are left with
\beq
\Box R_\abgd = 8 \pi G^\prime_N (\prt_{\be\de} {\tilde T}_{\al\ga}-\prt_{\be\ga} {\tilde T}_{\al\de}+\prt_{\al\ga} {\tilde T}_{\be\de}
-\prt_{\al\de} {\tilde T}_{\be\ga} ),
\label{identity3}
\eeq
where ${\tilde T}^\mn = T^\mn - (1/3) \et^\mn T^\al_{\pt\al\al}$ and there is a scaling
$G^\prime_N = \bar{\lambda} G_N$.  
Note that in standard GR, 
at this stage of the analysis, 
instead of the $1/3$ factor in ${\tilde T}^\mn$, 
there is a $1/2$.
This indicates a critical difference with GR, 
which appears to be unsuppressed by any small coefficient or parameter.
We exploit the resemblance of the result in \rf{identity3} to GR by adding and subtracting terms to the right-hand side, 
namely $0=(1/2)(\et_\mn - \et_\mn) T^\al_{\pt\al\al}$.
Rearranging, 
we obtain the GR result plus a correction.
Thus we have
\beq
\Box R_\abgd = (\Box R_\abgd )_{GR} 
+ \frac 43 \pi G^\prime_N 
( \et_{\al\ga}\prt_{\be\de}
-\et_{\al\de}\prt_{\be\ga} 
+\et_{\be\de}\prt_{\al\ga} 
-\et_{\be\ga}\prt_{\al\de}  ) 
 T^\mu_{\pt\mu\mu},
\label{identity4}
\eeq
where we note that $G_N^\prime$ is used also in the GR piece.
We specialize now to the six measurable components $R_{0j0k}$ in the gravitational wave measurements scenario.
This reduces Eq.~\rf{identity4} to
\beq
\Box R_{0j0k} = (\Box R_{0j0k} )_{GR} 
+ \frac 43 \pi G^\prime_N 
(-\prt_{jk}
+\de_{jk} \prt_0^2  
) ( T^{kk}-T^{00} ).
\label{identity5}
\eeq
At this stage we will use the standard retarded wave solution, 
and the result can be written as
\beq
R_{0j0k} = (R_{0j0k})_{GR} 
+\frac 13 G^\prime_N
(\prt_{jk}
-\de_{jk} \prt_0^2) 
\int d^3 r^\prime 
\frac{ [ T^{kk}-T^{00} ] (t- R, \vec r^\prime)}{R}.
\label{curvature}
\eeq
 Next we expand the solution for the wave zone assuming the source $T^\mn$ has compact support. 
 This computation makes use of known methods described in detail elsewhere \cite{pw14,Bailey:2023lzy,Nilsson:2023szw}.
 In the ``far away" wave zone 
 we seek only the $1/r$ contributions. 
 The leading multipole terms can be manipulated to the form,
\beq
\boxed{%
R_{0j0k} 
= -\frac {G_N^\prime}{r} \prt_t^4 (I_{jk})_{TT}
+\frac {G^\prime_N}{6 r} (\de_{jk}-n_j n_k)
\prt_t^4 (n_l n_m I^{lm} -I^{ll}),
\label{curvature2}%
}
\eeq
where the time dependence is $\ta=t-r$, 
and $n^j$ is a unit vector in the direction from the source to the field point.\footnote{We do not expect that the scaling of $G_N$ is observable.}

The first term in Eq.~\rf{curvature} is the result of the transverse traceless piece in GR, 
with two degrees of freedom
and the usual two polarizations.
The second term is transverse in the far away wave zone, 
but has an overall term proportional to the projection of the inertia tensor $I^{ij}$ that is perpendicular to $n^j$, 
and is hence a {\it breathing mode} \cite{Will:2018bme,Wagle:2019mdq}.
The latter represents an extra propagating degree of freedom, 
arising from the explicit symmetry breaking term in the action.

A peculiar thing about this result is that the strength of the extra term is of order unity relative to the GR term, and there is no coefficient or parameter multiplying the extra term that could be constrained.
This result therefore shows a ``discontinuity" in the solutions to this particular subset of the framework \rf{FE}, 
since there is no continuous limit that matches GR.
The same thing happens in the well-known massive gravity vDVZ discontinuity \cite{Hinterbichler:2011tt}.
Instead, 
the lack of observation of a scalar (trace) mode in the observed strain in GW detectors could be used to rule out this model, 
a subset of the EFT in the explicit breaking limit, 
outright!

\section{Linearized limit results}
\label{sec:genlin}

In this section we explore the consequences of explicit spacetime-symmetry breaking in a context where the constraints implied by the Bianchi identities as well as the identity of the degrees of freedom can be established from field equations linear in the metric fluctuations.
To keep the scope reasonable we restrict attention to the couplings in the action \rf{genAct}
that only involve $s_\mn$, 
thus only $e_2$ and $e_4$ are nonzero.
The metric will be expanded around a flat background as
\beq
g_\mn = \et_\mn + \ep h_\mn,
\label{metric}
\eeq
where $\ep$ is used to extract equations order by order in $h_\mn$ \cite{seifert09}.
By the same token we also expand the coefficient fields $s_\mn$:
\beq
s_\mn = \sb_\mn + \ep \tilde{s}_\mn, \quad {\rm (Explicit \, breaking) }
\label{sexp}
\eeq
where the constant background coefficients $\sb_\mn$ and the fluctuations $\tilde{s}_\mn$ are considered as given, 
{\it a priori}, background quantities.
We are still considering explicit symmetry breaking,
despite the resemblance to the spontaneous symmetry breaking
scenario \cite{b21}, and we note that the arbitrary background coefficients $\bar{s}_{\mu\nu}$ are not assumed small quantities.
The expansion \rf{sexp} can also be justified given that we are here assuming an asymptotically flat scenario where the coefficients relax to the constant values $\sb_\mn$ far from any matter sources. 

The linearized field equations (order $\ep$) can be obtained from Eq.~\rf{eq:Einstein}.
They can be written as 
\beq
\label{linearFE}
\begin{aligned}
&&G_\mn (1+e_2 \sb^\al_{~\al}) + \la_1[ \overline{s}_\mn R+  \sb^{\al\be} (\prt_\mu \prt_\nu - \et_\mn \Box) h_{\al\be} ]
+ e_4 \sb^{\al\be} {\cal G}_{\al\mu\nu\be} \\
&&= \ka (T_M)_\mn + \tfrac 12 e_4 \widetilde{G}_\mn + \la_1 (\prt_\mu \prt_\nu - \et_\mn \Box) \tilde{s}^\al_{\pt{\al}\al}, 
\end{aligned}
\eeq
where $\widetilde{G}_{\mu\nu}$ is the "Einstein tensor" for the fluctuations $\widetilde{s}_{\mu\nu}$, i.e. the canonical definition for $G_{\mu\nu}$ with the identification $g_{\mu\nu}\to\widetilde{s}_{\mu\nu}$, and $\mathcal{G}_{\alpha\mu\nu\beta}$ is the double dual of the linearized Riemann tensor \cite{b21}. In the above expressions, quantities like $\sb^{\al\be}$ are understood now as raised and lowered with the Minkowski metric $\et_\mn$, 
and again $\la_1=e_2+\tfrac 12 e_4$.

It is important to highlight symmetry properties of the equations at this linearized level, 
complementing the discussion in Section \ref{sec:minimaltrace}:
under a ``particle" diffeomorphism transformation, 
where the metric fluctuations transform but the coefficient fluctuations do not, 
the field equations \rf{linearFE} are not invariant (this is the physical diffeomorphism breaking).
Specifically, 
this is the transformation parametrized by $\xi_\mu$:
\beq
\bal
h_\mn &\rightarrow h_\mn + \prt_\mu \xi_\nu + \prt_\nu \xi_\mu,\\
\stw_\mn &\rightarrow \stw_\mn,
\eal
\label{particle}
\eeq
where we note that we take the background values $\eta_\mn$ and $\sb_\mn$ to be fixed under this transformation.
The breaking of the particle diffeomorphism symmetry by the particular case of \rf{linearFE} is a partial one,
any transformation $\xi_\mu$ for which 
\beq
\sb^\mn \prt_\mu \xi_\nu = 0 {\rm \,\,\,(residual \, diffeomorphisms)}
\label{residual}
\eeq
will still be a symmetry;
which leaves $3$ possible remaining unbroken symmetries.

In contrast to \rf{particle}, 
the general coordinate invariance of physics is maintained in 
that equations \rf{linearFE} {\it are} invariant 
under the general coordinate transformations parametrized by $\xi_\mu$
(an ``observer" transformation).
This transformation is given by,
\beq
\bal
h_\mn &= h^\prime_\mn + \prt_\mu \xi_\nu + \prt_\nu \xi_\mu,\\
\stw_\mn &= \stw^\prime_\mn + \sb^\al_{~\mu} \prt_\nu \xi_\al + \sb^\al_{~\nu} \prt_\mu \xi_\al.
\eal
\label{GCT}
\eeq
Both types of transformations are here taken to first order in small quantities $h_\mn$ and $\xi_\mu$, and $\stw_\mn$ \cite{k04,bk05,bk06,kl21,b21}, 
for consistency with the linearized approximation.
Note also that the role of the $4$ constraints stemming from the Bianchi identities \rf{BI1}, 
are contained in the partial divergence of the field equations, 
which yields $(e_2 + \tfrac 12 e_4)\prt^\mu (\sb_\mn R)=0$.
It should be noted that the framework used here does not include any mass-type terms for the fluctuations $h_\mn$, 
as considered in massive gravity approaches which also generically break diffeomorphism symmetry explicitly
\cite{PhysRevD.82.044020,Kostelecky:2021xhb}.

Modulo the extra terms with the coefficient fluctuations $\stw_\mn$, 
the field equations \rf{linearFE} are a subset of the general treatment of gauge-symmetry breaking terms in 
Ref.\ \cite{km18}.
The detailed-balance case of $\la_1=0$ for Eq.~\rf{linearFE} yields the linearized field equations 
of past publications \cite{bk06, km16, Bailey:2023lzy}, 
where solutions are already worked out (up to the piece $\widetilde{G}_\mn$), 
and tested in many experiments and observations.
Also, 
the case of 
$\la_1=0$ eliminates the explicit diffeomorphism violation entirely at this linearized level, 
which is clear upon examination of \rf{linearFE} under the transformation \rf{particle}.
Our focus here, 
however,
is on the effects when $\la_1 \neq 0$.

We highlight here an interesting occurrence, 
atypical of flat spacetime EFT approaches with background coefficients:
the entire right-hand side of \rf{linearFE} acts
as an effective energy-momentum tensor for the gravitational fluctuations $h_\mn$.
Thus, we are inclined to define, 
similar to the approach in Ref.\ \cite{bk06}, 
\beq
\Si_\mn = \ka (T_M)_\mn + \tfrac 12 e_4 \widetilde{G}_\mn + \la_1 (\prt_\mu \prt_\nu - \et_\mn \Box) \tilde{s}^\al_{\pt{\al}\al},
\label{sigma}
\eeq
where $\Si_\mn$ contains the matter energy-momentum tensor plus an independently conserved piece of the energy momentum tensor.
The latter piece depends entirely on the unknown background coefficient fluctuations $\stw_\mn$.
In the next subsection, 
we solve one special case where the effects
of these extra terms are included.

\subsection{Initial value solution for an explicit breaking case}
\label{A position space solution}

To illustrate some basic features we analyze a special case of \rf{linearFE}, 
where the field equations are tractable.
The case we consider shows the dependence of the solutions on the explicit breaking background choice.
First, 
we restrict attention to the case where only $\sb_{00}$ and $\stw^\prime_{00}$ are nonzero 
in a starting coordinate system $\{ x^{\prime \mu} \}$ and we set $e_4=0$.
This choice on $s_\mn$ represents a choice of symmetry-breaking background \cite{abn21,kl21}, with other choices of background possibly yielding different physical results.
For the spatial piece of the metric fluctuations we separate into traceless and trace pieces as
\beq
h_{ij}= (h^T)_{ij} - 2 \de_{ij} \Ps, 
\label{hijdecomp}
\eeq
where $(h^{T})_{ij}=h_{ij}-\tfrac 13 \de_{ij} h_{kk}$, 
and $\Ps = -\tfrac 16 h_{kk}$. 
We then use the general coordinate freedom \rf{GCT} to choose coordinates where the following conditions hold on the metric fluctuations:
\beq
\bal
h_{00} &= 0,
\\
\prt^i (h^{T})_{ij}&= 0.
\eal
\label{hybrid}
\eeq
This coordinate choice is a hybrid of the synchronous gauge 
and the transverse gauge \cite{Carroll:2004st};
this choice makes the analysis that follows simpler, 
and clarifies the degrees of freedom in the framework.
While this coordinate transformation eliminates $h_{00}$, 
it also shifts the background
$\stw^\prime_{00}$ by the undetermined function $\sb_{00}h^\prime_{00}$, 
according to Eq.~\rf{GCT}.
In addition, 
the background acquires the term $\stw_{0j} = -\sb_{00}\prt_j \xi_0$, but this latter change does not affect the field equations since only the combination $\stw^\al_{~\al}$ appears.

With the above choices, 
the field equations then take the form,
\beq
G_\mn (1-e_2 \sb_{00}) + e_2 \sb_\mn R 
+ e_2 P_\mn ({\tilde s}^\prime_{00}+\sb_{00} h^\prime_{00}) = \ka (T_M)_\mn ,
\label{SC1}
\eeq
where $P_\mn = \prt_\mu \prt_\nu - \et_\mn \Box$. 
Note that we {\bf do not} absorb $h^\prime_{00}$ into $\stw^\prime_{00}$; this is so that
we do not affect the {\it a priori} nature of the given background function $\stw^\prime_{00}$.
Instead, $h^\prime_{00}$ becomes a function that is determined dynamically below.
This means that we have {\it not} eliminated four field components completely with \rf{hybrid}, 
since one undetermined function remains.
If one takes the alternative route of calling the combination
${\tilde s}^\prime_{00}+\sb_{00} h^\prime_{00}$ the new {\it a priori} given background, 
then one is choosing a different symmetry-breaking background, 
and further, 
as we show below, 
one immediately runs into issues with constraints directly on the matter stress-energy tensor.

Next we break the field equations into space and time components: 
\beq
\bal
 \nabla^2 [ 2 (1+e_2 \sb_{00})\Ps -e_2 \Om ] 
- 6 e_2 \sb_{00} \ddot{\Ps} -2 e_2 \sb_{00} \prt_i h_{0i}&= \ka T^{00},\\
-\tfrac 12 (\prt_i \prt_j h_{0j} - \nabla^2 h_{0i}) 
- \prt_i \prt_0 ( 2 \Ps \de - e_2 \Om ) &= \ka T^{0i},\\
-\tfrac 12 \Box (h^{TT})_{ij} \de + \de_{ij} (\prt_k \dot{h}_{0k} \de + 2\ddot{\Ps} \de - e_2 \ddot{\Om})
- \tfrac 12 ( \prt_i \dot{h}_{0j} + \prt_j \dot{h}_{0i} ) 
+ \left( \prt_i \prt_j - \de_{ij} \nabla^2 \right) 
[ \Ps \de - e_2 \Om ]
&= \ka T^{ij},
\eal
\label{SC2}
\eeq
where $\de=1- e_2 \sb_{00}$, 
$\Om = \stw^\prime_{00}+\sb_{00} h^\prime_{00}$, 
and $\nabla^2 =  \de^{ij} \prt_i \prt_j$.
Note that we have labelled the spatial metric components as transverse and traceless,
$(h^{TT})_{ij}$, 
on account of equation \rf{hybrid}.
We solve these equations for the remaining ``unknowns" (since $h_{00}=0$ and $\prt^i (h^{T})_{ij}=0$), 
which are the combination $\Om = \stw^\prime_{00}+\sb_{00} h^\prime_{00}$, 
the three vector components $h_{0i}$, 
$\Ps$, 
and the transverse-traceless components $(h^{TT})_{ij}$.
This is {\it seven} functions, rather than the six in GR,
again because we are explicitly breaking the diffeomorphism symmetry.
The equations are solved below to the point where the number of {\it degrees of freedom}, 
i.e., the fields for which we can freely specify the initial value and first derivative, 
is clear \cite{Wald:1984rg,Diaz:2014yua}.
While the equations \rf{SC1} are really a special case of Eq.~\rf{FE3} in Section~\ref{sec:simplest}, 
this solution will make plainer the appearance of the degrees of freedom, 
using a step-by-step process below.
\vskip5mm
\noindent \underline{{\it Step 1.}} A spatial derivative of the second equation in \rf{SC2}, 
followed by imposing the conservation law $\prt_i T^{0i}=-\prt_0 T^{00}$,
yields the following initial value constraint equation,
\beq
\nabla^2 (2\Ps \de -e_2 \Om) = \ka T^{00},
\label{IVC1}
\eeq
which we take to fix $\Ps$ in terms of $\Om$ (ignoring homogeneous solutions to this equation).
Thus we have for $\Ps$,
\beq
\Ps  = \tfrac {e_2}{2\de} \Om -\tfrac{\ka}{2\de} \int d^3r^\prime \frac {T^{00} (t,\vec r^\prime )}{4\pi ||\vec r - \vec r^\prime||}.
\label{Psisoln}
\eeq
\vskip5mm
\noindent \underline{{\it Step 2.}}  
The trace of the last of equations \rf{SC2} yields
\beq
2 \prt_j \dot{h}_{0j}+3 \prt_0^2 (2\Ps \de -e_2 \Om)
-\nabla^2 (\Ps \de - e_2 \Om) = \ka T^i_{~i}.
\label{trace}
\eeq
Together, 
the first equation in \rf{SC2}, 
\rf{IVC1}, 
and \rf{trace} 
comprise three equations for three unknown functions
$\Om$, $\Ps$, and $\prt_j h_{0j}$.
To decouple them, 
we add $\de\times\big[$the first equation of \rf{SC2}$\big]$ to the product of $e_2 \sb_{00}$ and \rf{trace}, 
followed by substituting the constraint \rf{IVC1}, 
thereby obtaining the following equation for $\Om$:
\beq
e_2 \Box \Om = - \tfrac {\ka}{3} T^\al_{~\al}. 
\label{OmProp}
\eeq
This indicates that the combination $\Om$ is a propagating degree of freedom, 
determined for all $t$ (in the relevant domain)
once initial values and first time derivatives are given at an initial time $t_0$.
A formal solution can be written as
\beq
\Om = \tfrac {\ka}{3 e_2} 
\int d^3r^\prime \frac {T^\al_{~\al} (t_R,\vec r^\prime )}{4\pi ||\vec r - \vec r^\prime||},
\label{Omsoln}
\eeq
with the retarded time $t_R=t-||\vec r - \vec r^\prime||$.

This step is where, 
if we had re-interpreted $\Om$ as {\it the} background coefficient, 
{\it a priori} prescribed, 
we would have an immediate no-go constraint on matter where $T^\al_{~\al}$ is fixed by the background $\Om$.
Instead, 
with the arbitrary function $h^\prime_{00}$ remaining within $\Om$, 
we can treat $\Om$ as a dynamical function.
The result \rf{OmProp} also now completely determines $\Ps$ via \rf{Psisoln} (again we neglect homogeneous solutions).
In addition, 
the longitudinal component $\prt_j h_{0j}$
is also determined now, 
by equation \rf{trace}, 
specifically, 
\beq
\prt_i h_{0i} = \tfrac {\ka}{2\de} (T^{00} +T^{ii}) + \tfrac {3 \ka}{2\de} 
\int d^3r^\prime \frac {\ddot{T}^{00} (t,\vec r^\prime )}{4\pi ||\vec r - \vec r^\prime||}
-\tfrac {e_2}{2\de} \nabla^2 \Om.
\label{djh0jsoln}
\eeq
\vskip5mm
\vskip5mm
\noindent \underline{{\it Step 3.}} 
We return to the second of equations \rf{SC2} and use it to solve for the transverse
piece of the vector field $(h^T)_{0i}$; $\prt^i (h^T)_{0i}=0$.
The term in this equation involving the combination $2\Ps \de - e_2 \Om$ can be re-expressed in terms of $T^{00}$ using \rf{IVC1} or \rf{Psisoln}.
Imposing the conservation law for the matter stress-energy tensor $\prt_0 T^{00} +\prt_i T^{0i}=0$, 
and manipulating the result yields,
\beq
\bal
\nabla^2 h^T_{0i} &= 2 \frac {\ka}{\de} \left ( T^{0i} + \prt_i \int d^3r^\prime \frac {\prt^\prime_j T^{\prime 0j} }{4\pi ||\vec r - \vec r^\prime|| } \right)\\
&= 2 \frac {\ka}{\de} T_\perp^{0i},
\eal
\label{hT0ieqn}
\eeq
where the combination appearing on the right-hand side is the transverse piece of the matter current $T^{0i}$
(see, for example, the solution for the transverse vector potential in chapter 6.3 of Ref.\ \cite{Jackson}).
The solution is thus 
\beq
h^T_{0i}= - \tfrac {2\ka}{\de} 
\int d^3r^\prime \frac {T_\perp^{0i} (t,\vec r^\prime )}{4\pi ||\vec r - \vec r^\prime||}.
\label{hT0isoln}
\eeq
\vskip5mm
\noindent \underline{{\it Step 4.}}  Finally, 
we can simplify the equation for the transverse-traceless piece of the spatial metric,
$(h^{T})_{ij}$, 
which is the last of equations \rf{SC2}.
Insertion of the solutions $h_{0i}$, 
$\Ps$, and $\Om$ 
into the third equation in \rf{SC2}, 
after simplification, yields
\beq
\bal
-\frac 12 \Box (h^{TT})_{ij} &= \frac {\ka}{\de} \Big[ T^{ij} - \tfrac 13 \de^{ij} T^k_{~k} 
+ \prt_i \prt_k \int d^3r^\prime \frac {T^{\prime kj}}{4\pi ||\vec r - \vec r^\prime|| }
+ \prt_j \prt_k \int d^3r^\prime \frac {T^{\prime ki}}{4\pi ||\vec r - \vec r^\prime||}
-2 \prt_i \prt_j \int d^3r^\prime \frac{||\vec r - \vec r^\prime||}{8\pi} \prt^\prime_k \prt^\prime_l T^{\prime kl} \\
& -\frac 12 (\prt_i \prt_j - \tfrac 13 \de_{ij} \nabla^2)
\left( \int d^3r^\prime \frac {T^{\prime kk}}{4\pi ||\vec r - \vec r^\prime||} -3 \int d^3r^\prime \frac {||\vec r - \vec r^\prime||}{8\pi} \prt^\prime_k \prt^\prime_l T^{\prime kl}
\right) \Big],
\label{TT1}
\eal
\eeq
where we have used standard post-Newtonian ``super-potentials" defined in the literature \cite{Will:2018bme}
as well as matter conservation law identities like $\ddot{T}^{00} = \prt_j \prt_k T^{jk}$.
What remains in the brackets in \rf{TT1} is the transverse and traceless projection of the stress-energy tensor \cite{Ashtekar:2017wgq}.
The result is the standard wave equation for this part of the metric:
\beq
\bal
-\frac 12 \Box (h^{TT})_{ij} &= \frac{\ka}{\de}  (T_{TT})^{ij}. 
\label{TT2}
\eal
\eeq
This, of course, 
contains two degrees of freedom for the usual transverse-traceless modes of GR, 
and they can solved for as in \rf{Omsoln}.

To recap the above,
we solved the field equations \rf{SC1} for the metric fluctuations $h_\mn$, 
using a coordinate condition \rf{hybrid}.
There are $3$ (propagating) degrees of freedom:
$1$ in $\Om$, 
and two in $(h^{TT})_{ij}$.
The remaining fields $h_{0i}$ and $\Ps$ are fixed by initial value constraints.
As an additional check, 
with the solutions above, 
it can be shown that the linearized Ricci curvature $R=-2\prt^i h_{0i}+4\nabla^2 \Ps - 6 \ddot{\Ps}=0$, 
thus the solution satisfies
the Bianchi identity-implied constraints $(e_2 + \tfrac 12 e_4)\prt^\mu (\sb_\mn R)=0$.

To find the physical effects of our solutions, 
we can proceed as in section \ref{solution for the curvature in the wave zone}, 
using the same method to obtain an effective wave equation for the curvature.
Up to an unobservable scaling, 
the field equations are the same as \rf{FE3wf} of the previous section, 
due to the wave equation \rf{OmProp} being the same result as in \rf{wave1}.
Therefore the final result in the wave zone \rf{curvature2} applies and thus the only observable beyond-GR effects are from the mode $\Om$ in this context.
We summarize the results in Table~\ref{tab:DOFtable}, 
where we include all solved components, 
the equations for their solutions, 
whether they are degrees of freedom (D.O.F.), 
if they are propagating, 
and if they show up in the wave zone (WZ) curvature.
\begin{center}
\begin{table}[h!]
    \begin{tabular}{ c c c c c}
      Field \,\, & Solution \,\, & D.O.F. \,\, & Propagating \,\,& WZ curvature observable \\
      \hline
      $h_{00}$ & $0$ & N/A & N/A & N/A \\
      $\prt^i (h^T)_{ij}$ & $0$ & N/A & N/A & N/A \\
      $(h^{TT})_{ij}$ & Eq.\ \rf{TT2} & 2 & Yes & Yes\\
      $\Ps$ & Eq.\ \rf{Psisoln} & 0 & No & No \\
      $h_{0i}$ & Eqs.\ \rf{hT0isoln},\rf{djh0jsoln} & 0 & N/A & No \\
      $\Om$ & Eq.\ \rf{Omsoln} & 1 & Yes & Yes \\
      \hline
     \end{tabular}
     \caption{Character of the solutions for the metric fluctuations $h_\mn$ to the linearized field equations in \rf{SC1} for the special background choice with nonzero $s^\prime_{00}=\sb_{00}+\stw^\prime_{00}$ only.} 
     \label{tab:DOFtable}
     \end{table}
\end{center}

It is helpful at this point to summarize this subsection as follows: we solved the explicit diffeomorphism-symmetry breaking linearized field equations \rf{linearFE} for a special case \rf{SC1}; we made a particular choice of background and found a solution which has an initial value formulation \cite{Wald:1984rg}, is consistent with the Bianchi identity-implied constraints \rf{BI1}, contains one extra degree of freedom relative to linearized GR, and has measurable consequences via the curvature result in \rf{curvature2}.
We find these results consistent with Ref.\ \cite{abn21}, where, for a particular choice of background, 
extra degrees of freedom were also obtained using the Dirac-Hamiltonian analysis (but without assuming linearized gravity).
Nonetheless, the solution is likely ruled out by the nonobservation of (large) extra GW polarizations.

\subsection{Momentum space matrix solution}
\label{momentum space matrix solution}

The general coefficient case is considerably more complicated due to the anisotropic coefficients in $\sb_\mn$.
One straightforward, albeit tedious, 
method to explore the dynamical content of Eq.~\rf{linearFE}, 
is to use Fourier decomposition in momentum space, 
thereby reducing the problem to an algebraic one.
Indeed this has long been used for studies of propagation effects in spacetime-symmetry breaking \cite{Kostelecky:2000mm,km01,km02,Cheng:2006us,km09,Schreck:2011ai,km13,Nascimento:2021rlg}.
Thus for the metric fluctuations we use,
\beq
h_\mn (x) = \frac {1}{(2\pi)^4} \int d^4p \, \htw_\mn (p) e^{-ip_\mu x^\mu}
\label{mom},
\eeq
with a similar ansatz for $\tilde{s}_\mn$.
The tilde notation over fields indicates the Fourier transform in what follows.
Denoting the right-hand side of \rf{linearFE}
as $\Si_\mn$, 
as in \rf{sigma}, 
we can write the equation in a suggestive form
\beq
M_\mn^{\pt{\mn}\al\be} \htw_{\al\be} = \tilde{\Si}_\mn,
\label{matrix1}
\eeq
where now the ``source" $\tilde{\Si}_\mn$ is expressed in momentum space, and thus depends on $p_\mu$.
The quantities $M_\mn^{\pt{\mn}\al\be}$ are obtained by factoring out the metric fluctuations from the left-hand side of \rf{linearFE},
followed by the substitution $\prt_\mu \rightarrow -ip_\mu$
\cite{km09}, after which we obtain
\beq
M_\mn^{\pt{\mn}\al\be} =
-\frac 12 p^2 \de^{(\al}_{\pt{\al}\mu} \de^{\be)}_{\pt{\be}\nu} \cdots + \la_1 [ \sb_\mn (\et^{\al\be} p^2 - p^\al p^\be ) +  \sb^{\al\be} (p_\mu p_\nu - \et_\mn p^2)] + e_4 \sb^{\ga\de} ( \tfrac 12 p_\ga p_\mu \de^\al_\nu \de^\be_\ga + \cdots).
\label{matrix2}
\eeq

The system of algebraic equations in \rf{matrix1} can be recast into a $10\times10$ matrix equation for the ten components of the symmetric metric fluctuations $\htw_\mn = \htw_{\nu\mu}$.
We use a $10$ component array with ordering ${\bf \htw}= (\htw_{00}, \htw_{01},\htw_{02},\htw_{03}, \htw_{11}, \htw_{12}, \htw_{13},\htw_{22},\htw_{23},\htw_{33})$ 
and a corresponding array for $\Si_\mn$ (${\bf \Si}= ({\tilde \Si}_{00}, {\tilde \Si}_{01},...)$).
The system of equations then takes the form
\beq
{\bf M} \cdot {\bf \htw}={\bf \Si}.
\label{matrix3}
\eeq
With this set-up, 
determining
the properties of the matrix ${\bf M}$
will determine the solutions in momentum space and  
aid in finding the independent degrees of freedom.
Such methods have been employed previously in cases of spontaneous-symmetry breaking \cite{bk05},
and explicit symmetry breaking \cite{Kostelecky:2021xhb}.

\subsubsection{GR case}

To simplify the process, 
we first use rotational freedom to align the spatial momentum $p^j$ along the $3$ direction; thus for instance, $p^2 = p^\mu p_\mu = p_3^2-p_0^2$.
It is instructive to consider the general relativity case first,
where the matrix ${\bf M}$ takes the form, 
\beq
M_{GR}=
\left(
\begin{array}{cccccccccc}
 0 & 0 & 0 & 0 & \frac 12 p_3^2  & 0 & 0 & \frac 12 p_3^2  & 0 & 0\\ 
 0 & -p_3^2  & 0 & 0 & 0 & 0 & p_0 p_3  & 0 & 0 & 0 \\ 
 0 & 0 & -p_3^2  & 0 & 0 & 0 & 0 & 0 & p_0 p_3 & 0 \\ 
 0 & 0 & 0 & 0 & - p_0 p_3  & 0 & 0 & - p_0 p_3  & 0 & 0 \\ 
 \frac 12 p_3^2 & 0 & 0 & - p_0 p_3  & 0 & 0 &0 & -\frac 12 p^2  &0 & \frac 12 p_0^2\\ 
 0 & 0 & 0 & 0 & 0 & p^2 & 0 & 0 & 0 \\ 
 0 &p_0 p_3 & 0 & 0 & 0 & 0 & -p_0^2  & 0 & 0 & 0 \\ 
 \frac 12 p_3^2  & 0 & 0 & -p_0 p_3  & -\frac 12 p^2 & 0 & 0 & 0 & 0 & \frac 12 p_0^2\\ 
 0 & 0 &p_0 p_3 & 0 & 0 & 0 & 0 & 0 & -p_0^2 & 0 \\ 
 0 & 0 & 0 & 0 & \frac 12 p_0^2 & 0 & 0 & \frac 12 p_0^2 & 0 & 0 \\ 
\end{array}
\right),
\label{Mgr}
\eeq
with the right-hand side of \rf{matrix1} being $\Si_\mn = \ka (T_M)_\mn$.
For the GR case, the matrix rank is $6$, leaving 4 of the 10 equations that cannot be used to solve for components of the metric.
For instance, 
using $6$ of the matrix equations to solve for the $6$ spatial components of the metric $\htw_{ij}$ yields the following column vector of solutions for ${\bf \tilde{h}}$:
\beq
{\bf \htw}_{GR}=
\left(
\begin{array}{c}
\htw_{00} \\
\htw_{01} \\
\htw_{02} \\
\htw_{03} \\
-\ka \left(\frac {(T_{11} - T_{22} )}{p^2}
+\frac {T_{00}}{p_3^2} \right) \\
-\frac {2\ka T_{12}}{p^2} \\ 
\frac {p_0 p_3 \htw_{01}-2\ka T_{13}}{p_0^2} \\
\ka \left(\frac {(T_{11} - T_{22} )}{p^2}
-\frac {T_{00}}{p_3^2} \right) \\ 
\frac {p_0 p_3 \htw_{02}-2\ka T_{23}}{p_0^2} \\
\frac {2 p_0 p_3 \htw_{03}-p_3^2 \htw_{00}+ \ka (T_{11}+T_{22} -T_{33}+T_{00})}{p_0^2}
\end{array}
\right),
\label{hGR}
\eeq
where, 
when needed, 
we have simplified expressions using the matter stress-energy conservation law
$p^\mu T_{\mu\nu}=0$.
The $\htw_{0\mu}$ components are gauge (for instance, one can choose a gauge $\htw_{0\mu}=0$).
One can clearly see the two propagating degrees of freedom in $\htw_{11}-\htw_{22}$ and $\htw_{12}$;
propagation being indicated by the presence of $p^2$ in the denominator.
The rest of the components are associated with non-propagating effects, 
but are part of the complete solution.

To complete the GR discussion we calculate the (gauge-independent) curvature tensor components $R_{0i0j}$, 
relevant for GW measurements.
In momentum space, 
the curvature components are the six quantities $R_{0i0j} = (1/2)(p_0^2 \htw_{ij} -p_0 p_i \htw_{0j} -p_0 p_j \htw_{0i}+p^2 \htw_\al^{~\al} ) $.
We display them here in terms of the plus, cross, breathing, longitudinal and vector components
for waves; recalling that we assumed orientation such that $p_\mu = (p_0,0,0,p_3)$.
Thus the transverse directions are
$1$ and $2$.
We find,
\beq
\bal
R_{0101}-R_{0202} &= -\frac {\ka p_0^2 (T_{11}-T_{22})}{p^2}, \\
R_{0102} &= \frac {\ka p_0^2 T_{12}}{p^2}, \\
R_{0101}+R_{0202} & = \ka  T_{33}, \\
R_{0303} &= \ka \tfrac 12 (T_{11} + T_{22} - T_{33} + T_{00}), \\
R_{0103} &= -\ka T_{13} , \\
R_{0203} &= -\ka T_{23}.
\eal
\label{GRcurv}
\eeq
Note that the undetermined $\htw_{0\mu}$ components completely vanish from the curvature, 
as expected.
The first two results confirm the standard GR plus and cross (propagating) polarizations. 
Additionally, 
the latter four curvature projections are all ``contact" terms that vanish outside the source (when the results are put back in position space).

\subsubsection{Explicit symmetry-breaking case}

To explain the matrix method for the explicit symmetry breaking case, 
and to display the form of the matrix ${\bf M}$, 
we begin by assuming for the moment only one nonzero background coefficient $\sb_{00}$ (thus $\sb^\al_{~\al}=-\sb_{00}$).
As before, 
we align $p^j$ along the $3$ direction to simplify the calculation. 
Also for convenience, 
we define the combination 
$\la_2=e_2-\tfrac 12 e_4$,
and as used earlier,
$\la_1$ and $\de=1-e_2\sb_{00}$.
In this case ${\bf M}$ takes the form,
\beq
M=
\left(
\begin{array}{cccccccccc}
 -2 \la_1 p_3^2 \sb_{00} & 0 & 0 & 2\la_1 p_0 p_3 \sb_{00} & P_1 
 & 0 & 0 & P_1 & 0 & -\la_1 p_0^2 \sb_{00} \\ 
 0 & -p_3^2 \de & 0 & 0 & 0 & 0 & p_0 p_3 \de  & 0 & 0 & 0 \\ 
 0 & 0 & -p_3^2 \de & 0 & 0 & 0 & 0 & 0 & p_0 p_3 \de  & 0 \\ 
 2 \la_1 p_0 p_3 \sb_{00} & 0 & 0 & 0 & - p_0 p_3 \de  & 0 & 0 & - p_0 p_3 \de  & 0 & 0 \\ 
P_1 & 0 & 0 & - p_0 p_3 \de & 0 & 0 &0 & P_2  &0 & \tfrac 12 p_0^2 \de  \\ 
 0 & 0 & 0 & 0 & 0 & -2 P_2 & 0 & 0 & 0 & 0\\ 
 0 & p_0 p_3 \de & 0 & 0 & 0 & 0 & -p_0^2 \de & 0 & 0 & 0 \\ 
 P_1 & 0 & 0 & -p_0 p_3 \de & P_2 & 0 & 0 & 0 & 0 & \tfrac 12 p_0^2 \de\\ 
 0 & 0 & p_0 p_3\de & 0 & 0 & 0 & 0 & 0 & -p_0^2 \de & 0 \\ 
 -\la_1 p_0^2 \sb_{00} & 0 & 0 & 0 & \tfrac 12 p_0^2 \de & 0 & 0 & \tfrac 12 p_0^2 \de & 0 & 0 \\ 
\end{array}
\right),
\label{M00}
\eeq
where we define the following dispersion quantities:
\beq
\bal
P_1 &= \frac 12 p_3^2 [ 1 + (e_2+e_4) \sb_{00} ] - \la_1 \sb_{00} p_0^2 \\
P_2 &= \frac 12 \left[ p_0^2 \de - p_3^2 (1 + (e_4-e_2) \sb_{00}) \right] 
\eal
\label{disp}
\eeq

We find that the matrix ${\bf M}$ for \rf{M00}
has rank $7$ provided $\la_1 \neq 0$.
Thus we can solve the matrix for $7$ of the metric fluctuation components of ${\bf \htw}$.
The remaining $3$ equations are constraints, 
indicating the leftover $3$ gauge degrees of the freedom, 
since the diffeomorphism symmetry is not completely broken.
We choose seven of the equations in \rf{matrix2} and solve for $\htw_{00}$ in addition to the spatial components $\htw_{ij}$.
The solutions are,
\beq
{\bf \htw}_{\sb_{00}}=
\left(
\begin{array}{c}
\frac {\de (\Sitw_{11}+\Sitw_{22}+\Sitw_{33})- (1-(e_2+e_4)\sb_{00})\Sitw_{00}}
{3 \la_1 \sb_{00} [p_3^2 (1- (e_2+e_4/3)\sb_{00}) -p_0^2 \de ]} \\
0 \\
0 \\
0 \\
\frac 23 \frac {-p_0^2 \de (2\Sitw_{11}-\Sitw_{22}-\Sitw_{33})
+ p_3^2 [(1+(e_4-e_2)\sb_{00})\Sitw_{00}
+2 \de \Sitw_{11} - (1-(e_4+e_2)\sb_{00}) \Sitw_{22}
-2 (1-\la_2 \sb_{00}) \Sitw_{33}]}
{[p_3^2 (1 + (e_4-e_2) \sb_{00})
-p_0^2 \de][p_3^2 (1 - (2e_2/3+ e_4/2) \sb_{00}) - p^2_0 \de ] }
 \\
\frac {2 \Sitw_{12}}{p_3^2 (1 + (e_4-e_2) \sb_{00})-p_0^2 \de }  \\ 
-\frac {2 \Sitw_{13}}{p_0^2 \de}  \\
\frac 23 \frac {p_0^2 \de (\Sitw_{11}-2\Sitw_{22}+\Sitw_{33})
+ p_3^2 [(1+(e_4-e_2)\sb_{00})\Sitw_{00}
 - (1-(e_4+e_2)\sb_{00}) \Sitw_{11} + 2 \de \Sitw_{22}
-2 (1-\la_2 \sb_{00}) \Sitw_{33}]}
{[p_3^2 (1 + (e_4-e_2) \sb_{00})
-p_0^2 \de][p_3^2 (1 - (2 e_2/3+ e_4/2) \sb_{00}) - p^2_0 \de ] } \\ 
-\frac {2 \Sitw_{23} }{p_0^2 \de} \\
\frac 13 \frac {p_3^2 [(1 +(3e_2 - e_4)\sb_{00})\Sitw_{00} - (1-(3e_2-e_4)\sb_{00})(\Sitw_{11} + \Sitw_{22}) ] - p_0^2 [(1 +(7e_2 + 4e_4)\sb_{00})\Sitw_{00}+ 2\la_1 \sb_{00} (\Sitw_{11} + \Sitw_{22} - 2 \Sitw_{33}) ]  }
{ \la_1 p_0^2 \sb_{00} [ p_3^2 ( 1 - (e_2 - e_4/3) \sb_{00}) - p_0^2 \de ] }, 
\end{array}
\right),
\label{hs00}
\eeq
where we have set $\htw_{01}=0,\htw_{02}=0,\htw_{03}=0$ using the coordinate freedom.
At this point it appears the solutions support multiple speeds of waves given the differing propagators appearing in several of the components.  
Also, 
$\htw_{00}$ appears to be a propagating mode, 
unlike in GR. We note that it is not immediately clear that the effective energy density in Eq.~(\ref{linearFE}) is bounded from below. For the case in Eq.~(\ref{linearFE}) as well as for the results in the present section, we would need to construct the properly time-averaged stress-energy pseudotensor to ascertain whether the extra propagating modes found above and later in this section are ghosts or not; however, this lies beyond the scope of the present work.

To clarify the physicality of the results, we next we find the curvature components, 
which are more illuminating since they are coordinate independent 
at the linearized gravity level, 
and they correspond directly to observables.
The components of curvature for this case ($\sb_{00} \neq 0$ only) are given by
\beq
\bal
R_{0101}-R_{0202} &= \frac {p_0^2 (\Sitw_{11}-\Sitw_{22})}
{p_3^2(1 +(e_4-e_2) \sb_{00})-p_0^2 \de }, \\
R_{0102} &= \frac {p_0^2 \Sitw_{12}}{p_3^2(1 +(e_4-e_2) \sb_{00})-p_0^2 \de } , \\
R_{0101}+R_{0202} & = \frac { p_0^2 (2\Sitw_{00} - 2\Sitw_{33} +\Sitw_{11} + \Sitw_{22}) }
{3 [p_3^2 (1 -(e_2-\tfrac 13 e_4) \sb_{00}) - p_0^2 \de ] }, \\
R_{0303} &=\frac { (p_3^2- p_0^2) (2\Sitw_{00} - 2\Sitw_{33} +\Sitw_{11} + \Sitw_{22}) }
{3 [p_3^2 (1 - (e_2-\tfrac 13 e_4)\sb_{00}) - p_0^2 \de ]}, \\
R_{0103} &= - \frac{1}{\de} \Sitw_{13} , \\
R_{0203} &= -\frac{1}{\de} \Sitw_{23}.
\eal
\label{s00curv}
\eeq
These results tell us the following:
firstly, 
there appear to be four propagating curvature modes here; the two standard plus and cross modes and two beyond-GR modes: 
a breathing mode ($R_{0101}+R_{0202}$), 
and a ``totally" longitudinal mode ($R_{0303}$). 
Since it depends on the same combination of the source tensor $\Sitw_\mn$, 
the latter mode is not an independent degree of freedom.
Secondly,  
modified dispersion relations appear for the four modes; using $\om=p^0 (\vec p )$, 
they can be read off as
\beq
\bal
\om_{+,\times} &=p_3 \sqrt{ \frac { 1+(e_4-e_2) \sb_{00} }
{1-e_2\sb_{00} } },\\
\om_{S,L} &=p_3 \sqrt{\frac {1-(e_2-\tfrac 13 e_4) \sb_{00} }{1-e_2\sb_{00}}}.
\eal
\label{moddisp}
\eeq
The plus and cross modes travel with one speed while the breathing and longitudal modes travel with a different speed, 
a kind of pair-wise lateral birefringence.
Note that depending on the values of the parameters in \rf{moddisp} and the coefficient $\bar{s}_{00}$, 
superluminal or subluminal propagation can be obtained here;
we do not investigate this behavior and leave it for future work.
The extra polarizations
are unsuppressed by any coefficients, 
so would be ruled out if the nonobservation of large extra polarizations can be confirmed below a certain level.
Note that, 
due to the appearance of $\la_1$ in the denominator in Eq.\ \rf{hs00}, 
we cannot take the limit of these results of $\la_1 \rightarrow 0$;
instead we must solve this case separately (which was done in Ref.\ \cite{Bailey:2023lzy}).

For the results just obtained, 
we can also verify that they agree with the results previous subsection \ref{A position space solution}.
The latter results used coordinates where $h_{00}=0$ and $\prt_i (h^T)_{ij}=0$, 
which differ from the matrix method so we compare the curvature results.
In the limit $e_4=0$, 
we find that the purely longitudinal curvature $R_{0303}$ in \rf{s00curv}
indeed becomes a contact term proportional to $\Si_\mn$, 
in agreement with subsection \ref{A position space solution} and thus the result \rf{curvature2}.
Further the plus and cross curvature components take a standard form up to a scaling by $\de$ and the breathing mode propagates with the standard $p_3^2-p_0^2$, 
consistent with \rf{curvature2}.

In the general case, 
without making any assumptions on the coefficients,
when at least one coefficient has
a $0$ or $3$ component, 
the $10 \times 10$ matrix ${\bf M}$ has rank $7$
provided $\la_1 \neq 0$.
When $\la_1 = 0$, 
the rank reduces to $6$ for all coefficients.
Also, 
when $\la_1=e_2 + \tfrac 12 e_4=0$, 
the field equations in \rf{linearFE}
reduce to the special linearized limit
of the mass dimension $4$, 
gauge invariant EFT expansion \cite{bk06,km16}, 
which is a case consistent with an origin 
in spontaneous-symmetry breaking.
Note that for this latter case, 
the wave generation solutions, 
in the lowest post-Newtonian order, 
have already been obtained
in Refs.\ \cite{Bailey:2023lzy,Nilsson:2023szw}.

The momentum space solutions for $\tilde{h}_\mn$ 
in the general case with arbitrary $\la_1$ and $\la_2$,
and arbitrary coefficients
are too messy for concise presentation here; however, 
we display the solutions for the curvature
for different coefficients, 
using a ``one-at-a-time" method, similar to that used in data analysis \cite{Abbott_2017}.
The results on the matrix rank 
and propagating degrees of freedom 
in the curvature are detailed in Table~\ref{tab:DOFtable2}.
Also included are the ``prefactors" that occur in front of the signal for the extra polarizations - this gives a notion of the relative strength to the GR signal
(e.g., see \rf{s00curv}).
As already noted above, 
the signals can be unsuppressed by any ``parameter" in the framework.
For additional clarity, 
we have included the matrix, solution for the metric and the curvature projections for another nonzero coefficient case with an anisotropic coefficient in Appendix~\ref{app:fullmatrixsol}.

\begin{center}
\begin{table}[h!]
    \begin{tabular}{ l c c l l}
      Coefficient \,\, & Rank of $M$ \,\, & D.O.F. \,\,& WZ curvature modes (prop) & Prefactors \\
      \hline
      $\sb_{00}$ & $7$ & $3$ & $+,\times,S,L\star$ & $\tfrac 13$, $\tfrac 19 \sb_{00}$ \\
      $\sb_{01}$ & $7$ & $3$ & $+,\times,S,V_1\star,V_2\star$ & $\tfrac 13$, $\tfrac 13 e_4 \sb_{01}$, $ e_4 \sb_{01}$  \\
      $\sb_{02}$ & $7$ & $3$ & $+,\times,S,V_1\star,V_2\star$ & $\tfrac 13$, $e_4 \sb_{02}$, $ \tfrac 23 e_4 \sb_{02}$ \\
      $\sb_{03}$ & $7$ & $3$ & $+,\times,S, L\star$ & $\tfrac 13,  \tfrac 29 \sb_{03} $ \\
      $\sb_{11}-\sb_{22}$ & $6$ & $2$ & $+,\times,S\star$ &  $ e_2 (\sb_{11}-\sb_{22})$\\
      $\sb_{12}$ & $6$ & $2$ & $+,\times,S\star$ & $4 e_2 \sb_{12}$ \\
      $\sb_{13}$ & $7$ & $3$ & $+,\times,S,V_1\star,V_2\star$ & $\tfrac 13$, $\tfrac 13 e_4  \sb_{13}$, $e_4 \sb_{13}$ \\
      $\sb_{23}$ & $7$ & $3$ & $+,\times,S,V_1\star,V_2\star$ & $\tfrac 13$,$2e_2 \sb_{23}$, $\tfrac 43 e_4 \sb_{23}$ \\
      $\sb_{11}+\sb_{22}-2 \sb_{33}$ & $7$ & $3$ & $+,\times,S,L\star$ & $\tfrac 13$, $\tfrac {2}{27} e_4 (\sb_{11}+\sb_{22}-2 \sb_{33})$ \\
      $\sb^\mu_{~\mu}$ & $7$ & $3$ & $+,\times,S$ & $\tfrac 13$ \\
      \hline
     \end{tabular}
     \caption{Character of the solutions for the matrix equation \rf{matrix3}, with the coefficients considered nonzero one-at-a-time.
     We assume here that $\la_1 \neq 0$. The fourth column specifies whether the polarizations are plus ($+$), cross ($\times$), scalar ($S$), longitudinal ($L$), or vector ($V_1$, $V_2$), where the index $1,2$ indicates the basis vector projection.
     The star ($\star$) indicates that while the extra mode appears, 
     it is degenerate with the other modes (see \cite{Bailey:2023lzy}).
     The last column displays the prefactor for the beyond-GR polarizations; that is the factor in front of the combinations of $\Sitw_\mn$ appearing in the curvature expression.} 
     \label{tab:DOFtable2}
     \end{table}
\end{center}

Some general features are revealed from this study.
First, there are extra degrees of freedom (DOF) relative to GR for all but two coefficients.
These are the coefficients $\sb_\mn$ that have no time components nor spatial components along the direction of $p^j$;
namely $\sb_{12}$ and $\sb_{11}-\sb_{22}$.
Evidently, 
to change the mode content one needs background coefficients with projections along the wave vector or the time direction.
Second, 
while there are several extra polarizations compared to GR in the curvature solutions,
they are not all independent of one another.
Finally, 
extra polarizations occur that are unsuppressed by a background coefficient or a constant from the framework that could be constrained by experiment, 
as occurred earlier in Section \ref{solution for the curvature in the wave zone}.
This last point is most crucial for understanding whether the subset of the framework \rf{genAct} studied can be ruled out by experiment. 

All of the results in Table \ref{tab:DOFtable2} leave unspecified the unknown, background-dependent terms in $\Si_\mn$ \rf{sigma}.
One possibility is to set them to zero by assuming $\stw_\mn=0$.
Then $8$ of the results of Table \ref{tab:DOFtable2} are no-go; the results with a pre-factor of $1/3$ for the scalar polarization
are ruled out by non-observation of scalar polarizations at current sensitivities (see discussion below).
On the other hand, 
one could use observational constraints to place limits on, 
for instance, 
the combination $(2\Si_{00} - 2\Si_{33} +\Si_{11} + \Si_{22})/3$ that occurs in 
the observable curvature via the GW strain.
For a given source, 
we could then reconstruct $(T_M)_\mn$, 
and this result would limit the size of the extra terms in $\Si_\mn$ and hence $\stw_\mn$;
this kind of program could be carried out in future work.

\section{Summary and discussion}\label{sec:summary}

In this paper we investigated the curvature couplings in an EFT description of explicit spacetime-symmetry breaking with the goal of determining if no-go constraints are avoidable in several scenarios.
We studied the action in \rf{genAct}, 
which includes terms with coefficient traces not previously considered separately. 
It is possible to countenance combinations of the terms in the action that avoid no-go constraints implied by the Bianchi identities under certain conditions, as seen in Eq.~\rf{BI2} and the ``detailed balance" choices described  thereafter.
In Section \ref{sec:simplest}, we examined a special subset of the generalized action including trace terms with $s^\al_{~\al}$ and $t^{\al\be}_{~~~\al\be}$.
We found an extra degree of freedom in Eq.~\ref{wave1}, 
and established a match to scalar-tensor theory in an experimentally ruled-out limit. 
Further, we obtained beyond-GR gravitational-wave signals in Eq.~\rf{curvature2} which show a discontinuity with GR in this subset of the EFT. 

In Section~\ref{sec:genlin} we studied the linearized limit
of the framework, 
including explicit diffeomorphism breaking terms in \rf{linearFE}.
For a special case, 
the field equations were solved as an initial-value problem in Section \ref{A position space solution}, 
revealing an extra degree of freedom compared to GR (see Table \ref{tab:DOFtable}). 
We generalized this case in Section~\ref{momentum space matrix solution} by utilizing a momentum space $10\times10$ matrix equation for the metric fluctuations 
and a one-at-a-time coefficient method for the EFT coefficients;
the results are summarized in Table~\ref{tab:DOFtable2}.
The main conclusions of this part of the work is that extra polarizations for GWs are predicted, 
some of which could be ruled out due to having order unity strength compared to GR (discontinuity).

One can look at the current status of limits on extra polarizations from GW measurements:
ground-based detectors including LIGO and Virgo use Bayesian analysis to study different polarization modes with CBC sources \cite{Abbott_2016,Abbott_2017,Abbott_2019}, and
to test for the six polarizations \cite{Isi_2017} that many models predict beyond GR, 
a network of five or more detectors is needed \cite{Chatziioannou_2012,Takeda_2018}.
Yet some studies find constraints on the GW vector and scalar mode polarizations through analysis of the stochastic GW background \cite{Callister_2017,Abbott_2018} and several works implement the null-stream method to test for alternative polarizations than those for GR \cite{Hagihara_2018,Hagihara_2020,w2021nstreambased,liang2024unrevealing}, 
some providing constraint values when having electromagnetic signals to help provide sky localization \cite{Hagihara_2019,Pang_2020}.
Additional constraints on scalar-tensor theories have been provided by pulsar binary systems \cite{w2014testing,Freire2012,Stairs2003} while continuous waves generated from pulsars also provide a means to search for constraints on different polarizations \cite{Abbott_2018,Isi_2015,Isi_2017,Isi_2020}.  
Pulsar timing of stochastic GW backgrounds provide further constraints \cite{Alves_2011,Niu_2019,O_Beirne_2019,Bo_tier_2020,Lee_2008}, as with the recent data release from the North American Nanohertz Observatory for Gravitational Waves (NANOGrav), 
from the International Pulsar Timing Array (IPTA) \cite{Arzoumanian_2020,Chen_2021,Arzoumanian_2021,Antoniadis:2022pcn,EPTA:2023fyk}, as well as the Chinese Pulsar Timing Array \cite{Xu:2023wog}.
Double white dwarfs are also considered as a source for finding constraints using the upcoming LISA detector \cite{Nilsson:2023szw} along with the proposed TianQin mission \cite{Xie_2022}.

The tightest ratio constraints for scalar-tensor mixed polarization has been given by Ref.\ \cite{takeda_2021} via Bayesian inference across pure tensor, vector and scalar polarization models. 
Different set of parameter values are found for each, including the correlated luminosity distance and inclination angle, 
yet Bayes factors comparing different polarizations are found.
From the binary black hole merger event GW170814, 
they found a Bayes factor of 3.636 when comparing tensor to scalar and a factor 2.775 when comparing tensor to vector models. 
For the binary neutron star merger event GW170817, 
the factors are 60.271 and 51.043 for the tensor-scalar and tensor-vector, respectively. 
The large favor for the tensor models may indicate that the prefactors of $\frac{1}{3}$ and $\frac{1}{6}$ for the added scalar and vector models in the previous sections (Table~\ref{tab:DOFtable2}) might be ruled out. 
In a later work, 
testing a scalar-tensor model \cite{Takeda_2022}, 
constraints on the ratio of a scalar amplitude to the tensor amplitude, $R_{ST}$, are found from GW170814 as $R_{ST}\lesssim 0.10$ and from GW170817 as  $R_{ST}\lesssim 0.034$.

Much more work can be done along the lines of investigation in this paper. 
What would be the effect of explicit breaking terms on the near field, post-Newtonian limit? This could, 
like the scalar-tensor case discussed in Section \ref{Match to scalar-tensor models}, 
provide another angle on the results to rule out subsets of the framework.
 Additionally, what is the nature of the possible ghost-like terms found in the dispersion analysis of Section \ref{sec:genlin}? Could one write down a subset of the explicit-breaking EFT action (perhaps going beyond the minimal case in Eq.~(\ref{genAct})) that evades the GW discontinuity in analogy with the Vainshtein mechanism?~\cite{Babichev:2013usa}.
What role do the $t_{\al\be\ga\de}$ (untraced) play with explicit breaking? 
This is relevant since these coefficients previously escaped analysis in the linearized gravity regime under the assumption of spontaneous-symmetry breaking origin \cite{bk06,Bonder:2015maa}, 
while they did show up in cosmological settings \cite{Bonder:2017dpb}.
In this work, 
we have only investigated explicit breaking terms in the minimal EFT, 
but many higher-order terms
in the nonminimal EFT have been countenanced \cite{bkx15,km16}, 
and explicit-breaking limits can be studied.
Another subtlety to be explored is the choice of background; 
we choose the coefficients like $\sb_\mn$ to be fixed in the covariant tensor position.  
It would be of interest to 
investigate other explicit breaking background choices, 
as the physical results may differ, 
as discussed in this and other works \cite{abn21,kl21}.

Finally we include a cautionary remark to the reader. 
Care is required in examining results for observables in this paper, 
like curvature expressions for gravitational waves,
because most previous theory and data analysis assumed no explicit breaking terms, 
so prior limits on $\sb_\mn$ may not apply
\cite{datatables}.
For example, 
limits on coefficients $\sb_\mn$ exist from
many tests \cite{Muller:2007es,Chung:2009rm,Hees:2015mga,Shao:2017bgz}.  
Lunar Laser ranging places limits on combinations of coefficients in $\sb_\mn$ at the level of $10^{-8}-10^{-11}$ \cite{Bourgoin:2016ynf}. 
However, 
we cannot use these results with explicit breaking terms, 
like for example, 
the curvature expressions \rf{s01curv},
because the phenomenology of the latter results have not been meshed with the EFT framework under the assumption of spontaneous breaking.
Overlap is quite likely, 
especially since terms in the curvature calculated in this paper resemble results in the EFT of Ref.\ \cite{Bailey:2023lzy}. 

\acknowledgments
For Q.G.B. and K.O.A., 
this work was supported by National Science Foundation grant number 2308602.  N.A.N. was financed by CNES and IBS under the project code IBS-R018-D3, and acknowledges support from PSL/Observatoire de Paris.
The authors thank M.\ Schreck and V.A.\ Kosteleck\'y for useful feedback on the manuscript.

\appendix
\section{Full expressions for the most general case}\label{app:fullexpr}
The terms $E_{(4)\hdots (6)}$ in Eq.~(\ref{eq:generalEE}) read
\begin{equation}
    \begin{aligned}
    E_{(4)}^\mn &= -R s^{\mu\nu}-2G^{\al (\mu}s^{\nu)}_{~\al}
    +\frac{1}{2}g^\mn G^{\al\be}s_{\al\be}
    +\frac{1}{4}g^\mn R s^{\al}_{~\al}
    +\nabla_\al \nabla^{(\mu}s^{\nu)\la}
    -\frac{1}{2}\nabla^2 s^\mn
    -\frac{1}{2} g^\mn \nabla_\al \nabla_\be s^{\al\be}, \\
    E_{(5)}^\mn &= -G_{\al\be} t^{\al\mu\be\nu}
    -\frac{3}{2}R t^{\mu\al\nu}_{~~~~\al}
    -2G^{\al(\mu}t^{\nu)\be}_{~~~\al\be}
    +\frac{1}{4}g^\mn R t^{\al\be}_{~~\al\be}
    +\frac{1}{2}g^\mn G^{\al\be} t_{\al\ga\be}^{~~~~\ga}
    +\nabla_\al \nabla^{(\mu}t^{\nu)\be\al}_{~~~~~\be}
    -\frac{1}{2}\nabla^2 t^{\mu\al\nu}_{~~~~\al}, \\ 
    &-\frac{1}{2}g^\mn \nabla_\al \nabla_\be t_\ga^{~\al\ga\be} \\
    E_{(6)}^\mn &= \tfrac{1}{2} \bigl(- 3 R^{ \al\be\ga\mu} t_{\al\be\ga}^{~~~~\nu} 
    - 3 R^{ \al\be\ga\nu} t_{\al\be\ga}^{~~~~\mu} 
    + g^\mn R^{\al\be\ga\de} t_{\al\be\ga\de} 
    - 2 \nabla_\al \nabla_\be t^{\al \mu \be \nu}
    - 2 \nabla_\al \nabla_\be t^{\al \nu \be \mu}
    \bigr).
    \end{aligned}
\end{equation}
The corresponding contributions to the covariant divergence of the field equations can be written as
\begin{equation}
    \begin{aligned}
        \nabla_\mu E_{(4)}^{\mu\nu} &= \frac{1}{2}R^{\alpha\be}\nabla^\nu s_{\alpha\be}-\frac{1}{2}s^{\alpha\nu}\nabla_\alpha R-R^{\alpha\be}\nabla_\be s^{\nu}_{~\alpha}, \\
        \nabla_\mu E_{(5)}^{\mu\nu} &= 
        - \tfrac{1}{2} t^{\al \be \nu }{}_{\be } \nabla_{\al }R 
        + \tfrac{1}{2} R^{\alpha \be } \nabla^{\nu }t_{\al \ga \be }{}^{\ga} 
        -  R_{\al \be } \nabla_{\ga } t^{\ga \al \nu \be} 
        + t^{\al \be \nu \ga} \nabla_{\be} R_{\ga \al } 
        -  R^{\alpha \be } \nabla_{\al }
        t_{\be \ga }{}^{\nu \ga }, \\
        \nabla_\mu E_{(6)}^\mn &=
        \tfrac 12 R^{\al\be\ga\de} \nabla^\nu t_{\al\be\ga\de}
        -2 R^{\al\be\ga\de} \nabla_\de t_{\al\be\ga}^{~~~~\nu}
        +4 t_{\al\be\ga}^{~~~~\nu} \nabla^\al R^{\be\ga}.
        \label{DEterms}
    \end{aligned}
\end{equation}

\section{Matrix Method example with anisotropic coefficient}\label{app:fullmatrixsol}

Here we show the details from Table \ref{tab:DOFtable2}, 
for the particular case of a nonzero background coefficient
$\sb_{01}$.
The matrix in this case takes the form, 
\begin{scriptsize}
\beq
M=
\left(
\begin{array}{cccccccccc}
 0 & 2 \la_1 p_3^2 \sb_{01} & 0 & 0 & \tfrac 12 p_3^2 & 0 & 0 & \tfrac 12 p_3^2 & 0 & 0 \\ 
2 \la_1 p_3^2 \sb_{01} & -p_3^2 & 0 & -4\la_1 p_0 p_3 \sb_{01} & -2 \la_1 \sb_{01} p^2 & 0 & p_0 p_3  & 2 P^2 \sb_{01}& 0 & 2\la_1 p_0^2 \sb_{01} \\ 
 0 & 0 & -p_3^2 & 0 & 0 & -e_4 p_3^2 \sb_{01} & 0 & 0 & p_0 p_3  & 0 \\ 
0 & -4\la_1 p_0 p_3 \sb_{01}  & 0 & 0 & - p_0 p_3 & 0 & 0 & - p_0 p_3  & 0 & 0 \\ 
\tfrac 12 p_3^2 & -2\la_1 p^2 \sb_{01} & 0 & - p_0 p_3 & 0 & 0 &0 & -\tfrac 12 p^2 &0 & \tfrac 12 p_0^2  \\ 
 0 & 0 & -e_4 p_3^2 \sb_{01} & 0 & 0 & p^2 & 0 & 0 & e_4 p_0 p_3 \sb_{01} & 0\\ 
 0 & p_0 p_3 & 0 & 0 & 0 & 0 & -p_0^2 & -e_4 p_0 p_3 \sb_{01} & 0 & 0 \\ 
 \tfrac 12 p_3^2 & 2 P^2 \sb_{01} & 0 & -p_0 p_3 & -\tfrac 12 p^2 & 0 & -e_4 p_0 p_3 \sb_{01} & 0 & 0 & \tfrac 12 p_0^2 \\ 
 0 & 0 & p_0 p_3  & 0 & 0 & e_4 p_0 p_3 \sb_{01} & 0 & 0 & -p_0^2 & 0 \\ 
 0 & 2\la_1 p_0^2 \sb_{01} & 0 & 0 & \tfrac 12 p_0^2 & 0 & 0 & \tfrac 12 p_0^2  & 0 & 0 \\ 
\end{array}
\right),
\label{M01}
\eeq
\end{scriptsize}
where $p^2=p_3^2-p_0^2$
and $P^2 = \la_1 p_0^2 - e_2 p_3^2$.
The rank of this matrix is $7$, 
assuming $\la_1 \neq 0$.

We solve 7 of the matrix equations \rf{matrix3} 
for the metric components 
$\htw_{01},\htw_{11},\htw_{12},\htw_{13},\htw_{22},\htw_{23},\htw_{33}$, 
while, 
using residual gauge freedom, 
$\htw_{00}=0=\htw_{02}=\htw_{03}$.
The solutions are
\beq
{\bf \htw}_{\sb_{01}}=
\left(
\begin{array}{c}
0 \\
\frac {\Sitw^\al_{~\al} - 2 e_4 \sb_{01} \Sitw_{01}}{-6\la_1 \sb_{01} p^2} +O(\sb_{01}^2)\\
0 \\ 
0 \\
\frac 23 \frac {2 \Sitw_{11} - \Sitw_{22} - \Sitw_{33} +\Sitw_{00} + 2 e_4 \sb_{01} \Sitw_{01} +O(\sb_{01}^2) }{p^2 +O(\sb_{01}^2)} \\
\frac {2 (\Sitw_{12} + e_4 \sb_{01} \Sitw_{02})}{p^2 +O(\sb_{01}^2)}  \\ 
\frac {p_3 [ \Sitw^\al_{~\al} +(5\la_1+2 \la_2)\sb_{01}\Sitw_{01} - 6\la_1 \tfrac {p_0}{p_3} \sb_{01} \Sitw_{13}]}
{-6p_0 \la_1 p^2 \sb_{01} } \\
- \frac { 2 [\Sitw_{11} - 2 \Sitw_{22} + \Sitw_{33} - \Sitw_{00}  +4 e_2 \sb_{01} \Sitw_{01}]}{3 p^2 + O(\sb_{01}^2)} \\ 
 2 e_4 \sb_{01} \frac {p_3 \Sitw_{12}}{p_0 p^2} - \frac {2 \Sitw_{23}}{p_0^2} + O(\sb_{01}^2) \\
\frac 23 \frac {\Sitw_{11}+\Sitw_{22} +2 \Sitw_{00} - 2 \Sitw_{33} - 2 e_4 \sb_{01} \Sitw_{01} }{p_0^2} 
\end{array}
\right).
\label{hs01}
\eeq
Finally we use the solutions to calculate the $6$ curvature components $R_{0i0j}$.
We obtain
\beq
\bal
R_{0101}-R_{0202} &= \frac {p_0^2 (\Sitw_{11}-\Sitw_{22} +2 e_4 \sb_{01} \Sitw_{01})} 
{p^2 + O(\sb_{01}^2)}, \\
R_{0102} &= \frac {p_0^2 (\Sitw_{12} + e_4 \sb_{01} \Sitw_{02}) }{p^2 + O(\sb_{01}^2) } , \\
R_{0101}+R_{0202} & = \frac { p_0^2 (\Sitw_{11} + \Sitw_{22} -2 e_4 \sb_{01} \Sitw_{01} )}
{3 p^2 + O(\sb_{01}^2) } + \tfrac 23 \Sitw_{33}, \\
R_{0303} &= \frac 13 
\left( \Sitw_{11} + \Sitw_{22} - 2 \Sitw_{33} + 2 \Sitw_{00} - 2 e_4 \sb_{01} \Sitw_{01} \right), \\
R_{0103} &= -\Sitw_{13} - \tfrac 13 e_4 \sb_{01} \Sitw_{03} + \frac { e_4 \sb_{01}}{3p^2} p_0 p_3 (\Sitw_{11} - 2 \Sitw_{22} ) + O(\sb_{01}^2), \\
R_{0203} &= -\Sitw_{23} + \frac {e_4 \sb_{01} p_0 p_3 \Sitw_{12}}{p^2}  + O(\sb_{01}^2).
\eal
\label{s01curv}
\eeq
We conclude that, 
in addition to the usual plus and cross polarizations, 
the scalar, or breathing mode, is propagating.
The longitudinal mode is non-propagating, 
while the vector modes have propagating pieces.
The latter, however, 
don't appear to be independent of the other polarizations, 
and thus are not independent degrees of freedom.

\bibliography{refsv2}

\end{document}